\documentclass[floatfix,nofootinbib,twocolumn,showpacs,superscriptaddress, reprintnumbers,amssymb,letterpaper,amsmath,pra]{revtex4-2}

\usepackage{graphicx}
\usepackage{dcolumn}
\usepackage{bm}
\usepackage{xcolor}
\usepackage[colorlinks=true,allcolors=blue]{hyperref}
\usepackage{amsmath}
\usepackage{lineno}

\newcommand{\rmsemit}{\mbox{$\widetilde{\varepsilon}$}}
\newcommand{\mean}[1]{\mbox{$\langle{#1}\rangle$}}

\begin{document}


\title{Proposal for a Damping-Ring-Free Electron Injector for Future Linear Colliders}
\author{T. Xu}
\email{xu@niu.edu}
\affiliation{Northern Illinois Center for Accelerator \& Detector Development and Department of Physics, Northern Illinois University, DeKalb, IL 60115, USA} 
\author{M. Kuriki}
\affiliation{Hiroshima University, Higashi-hiroshima, Hiroshima, Japan 739-8527}
\author{P. Piot}
\affiliation{Northern Illinois Center for Accelerator \& Detector Development and Department of Physics, Northern Illinois University, DeKalb, IL 60115, USA} 
\affiliation{Argonne National Laboratory, Lemont, IL 60439, USA}
\author{J. G. Power}
\affiliation{Argonne National Laboratory, Lemont, IL 60439, USA}

\date{\today}

\date{\today}
\pacs{29.27.a, 41.75.Fr, 41.85.p}
\begin{abstract}
The current designs of future electron-positron linear colliders incorporate large and complex damping rings to produce asymmetric beams for beamstrahlung suppression. Here we present the design of an electron injector capable of delivering flat electron beams with phase-space partition comparable to the electron-beam parameters produced downstream of the damping ring in the proposed international linear collider (ILC) design. Our design does not employ a damping ring but is instead based on cross-plane phase-space-manipulation techniques. The performance of the proposed configuration, its sensitivity to jitter along with its impact on spin-polarization is investigated. The proposed paradigm could be adapted to other linear collider concepts under consideration and offers a path toward significant cost and complexity reduction. 
\end{abstract}

\maketitle

\section{Introduction \label{sec:intro}}

High-energy electron-positron ($e^-$/$e^+$) collisions have been invaluable engine of discovery in elementary-particle physics. TeV-class linear colliders (LC) will give access to energy-scale beyond the Standard Model~\cite{ellis-2001}. A critical metric to quantify the performances of an LC is the luminosity defined as 
\begin{eqnarray}
{\mathfrak L}=\frac{P_b}{E_b}\left(\frac{N}{4\pi\sigma_x^*\sigma_y^*}\right),
\end{eqnarray} 
where $N$ the single-bunch population, $E_b$ and $P_b$ respectively the energy and power associated with the beams and $\sigma^*_i$ refers to the horizontal ($i=x$) and vertical ($i=y$) beam sizes at the interaction point. During collision beam-beam interaction results in an envelop pinch which enhances luminosity while also resulting in an increase in energy spread due to beamstrahlung effects~\cite{dugan-2004-a}. A technique to mitigate beamstrahlung consists in using flat beams $\sigma_y \ll \sigma_x$~\cite{yokoya-2001-a}. In such a configuration the luminosity takes the form 
\begin{eqnarray} \label{eq:luminosity2}
{\mathfrak L}=\frac{P_b}{E_b}\frac{\sqrt{5}}{16 \alpha^2 \sqrt{3r_e}\pi} \frac{\sqrt{\gamma n_{\gamma}^3}} {\sqrt{\sigma_z}\sigma_y^*}
\end{eqnarray}
where $r_e$ is the classical radius of an electron, $\alpha\simeq 1/137$ the fine-structure constant, $n_{\gamma}$ the number of photon emitted via beamstrahlung,  $\gamma$ the Lorentz factor, and $\sigma_z$ is the bunch length. The required transversely asymmetric beams are naturally produced using damping rings (DRs) which generate a beam with asymmetric transverse normalized emittance partition $(\varepsilon_x, \varepsilon_y)$. Table~\ref{tab:lcparam} summarizes typical beam parameters achieved in design associated with few LC technologies. The latter table indicates that the required 6D phase-space brightness ${\cal B}_6\equiv Q/( \varepsilon_x \varepsilon_y \varepsilon_z)$ is $\sim 2$ orders of magnitude smaller than those achieved in state-of-the-art radiofrequency (RF) photoinjectors~\cite{krasilnikov-2012-a}. Such a feature was first recognized in Ref.~\cite{brinkmann-2001-a} where a linear transformation exploiting initial cross-plane correlation was proposed as a path to producing flat beams ($\varepsilon_y\ll \varepsilon_x$) using a photoinjector, i.e. without the need for a DR. In this latter work the achievable emittance ratio $\varrho\equiv \varepsilon_x/\varepsilon_y$ was comparable to the ones needed for ILC albeit at a much lower charge (0.5~nC in Ref.~\cite{brinkmann-2001-a} versus the required 3.2~nC~\cite{phinney-2007-a}). 

\begin{table}[hhhh!!]
\caption{Comparison of beam-parameter requirements for two conventional LC designs with parameter achieved in an RF photoinjector. The longitudinal emittance is evaluated as $\varepsilon_z\simeq\gamma \sigma_z\sigma_{\delta}$. The RF photoinjector used as an example is based on the L-band RF gun of the European X-ray FEL.  \label{tab:lcparam}}
\begin{center}
\begin{tabular}{l c c c}\hline\hline\
  & ILC & CLIC & RF gun\\
\hline
Reference &~\cite{phinney-2007-a} & ~\cite{clic-2018} & ~\cite{krasilnikov-2012-a} \\
Charge $Q$ (nC)  &  3.2  & 0.83 & 2     \\
Energy $E_b$ (GeV)  &  250  & 380 & $24\times 10^{-3}$     \\
$\varepsilon_{x}$ (\textmu{m}) &  10  & 0.9 & 1.3     \\
$\varepsilon_{y}$ (nm) &  35  & 20 & $1.3\times 10^3$      \\
$\sigma_{z}$ (mm) &  0.3  & 0.07 &  2.31    \\
$\sigma_{\delta}$ (\%) &  0.19  & 0.35 & $\sim 0.1$      \\
$\varepsilon_z$ (\text{m}) &  0.27  & 0.18 & $\sim 1.1\times 10^{-4}$     \\
${\cal B}_{6}$ (pC.\textmu{m}$^{-3}$) &  $3.4\times10^{-2}$  &  $0.25$ &  $\sim 11$    \\
\hline
\hline
\end{tabular}
\end{center}
\end{table}

In this paper we further expand the technique developed in~\cite{brinkmann-2001-a} by combining two cross-plane phase-space manipulations: a round-to-flat beam transformer (RFBT)~\cite{brinkmann-2001-a} followed by a transverse-to-longitudinal emittance exchanger (EEX)~\cite{cornacchia-2002-a,emma-2006-a}. These phase-space manipulations were developed and experimentally demonstrated over the last two decades~\cite{piot-2006-a,ruan-2011-a,sun-2010-a,groening-2014-a,ha-2016}. To illustrate the potential of the technique we consider the case of the ILC parameters and show that 6D brightness $\sim 2$ orders of magnitude larger than the nominal ILC injector can be attained in the proposed scheme. It should be noted that a similar approach employing cross-plane phase-space manipulations was proposed in a different parameter range to mitigate the micro-bunching instability in X-ray free-electron lasers (FELs)~\cite{emma-2006-a}. More generally, the idea of designing photoinjectors beamlines capable of producing tunable emittance partition via emittance repartitioning and emittance exchange was extensively discussed in Refs.~\cite{yampolsky-2010-a,carlsten-2011-a,duffy-2016-a}. Our approach confirms that emittance partition commensurate with requirements for an LC can be attained with a simple and compact ($<50$~m) beamline redistributing emittance typically produced in a conventional RF photoinjector.  

\section{theoretical background} 
\subsection{Transfer-matrix description of the concept}\label{ssec:linear}
In this section we describe the underlying principle of the proposed partitioning method. We introduce the coordinate of an electron as ${\boldsymbol{\mathcal Z}}^T=(x,x', y, y', z, \delta)$ where $(x,x')$ [resp. $(y,y')$] represents the position-angle coordinate associated to the horizontal [resp. vertical] phase space, $z$ is the longitudinal coordinate and $\delta$ its relative-momentum offset. All the coordinates are defined relative to a reference particle taken as the bunch barycenter. We further introduce the geometric beam emittance 
\begin{eqnarray}
\widetilde{\varepsilon}_i\equiv [\mean{{\cal Z}_i^2}\mean{{\cal Z}_{i+1}^2}- \mean{{\cal Z}_i {\cal Z}_{i+1}}^2]^{1/2}, 
\end{eqnarray}
for $i=1,3,5$ respectively corresponding to the horizontal $\rmsemit_x$, vertical $\rmsemit_y$, and longitudinal $\rmsemit_z$ geometric emittances. Additionally, the normalized emitance discussed in Sec.~\ref{sec:intro} are $\varepsilon_{\ell}\equiv \gamma \widetilde{\varepsilon}_{\ell}$ with $\ell=x,y, z$. 

\begin{figure}[hh!!!!!]
\centering
\includegraphics[width=0.90\columnwidth]{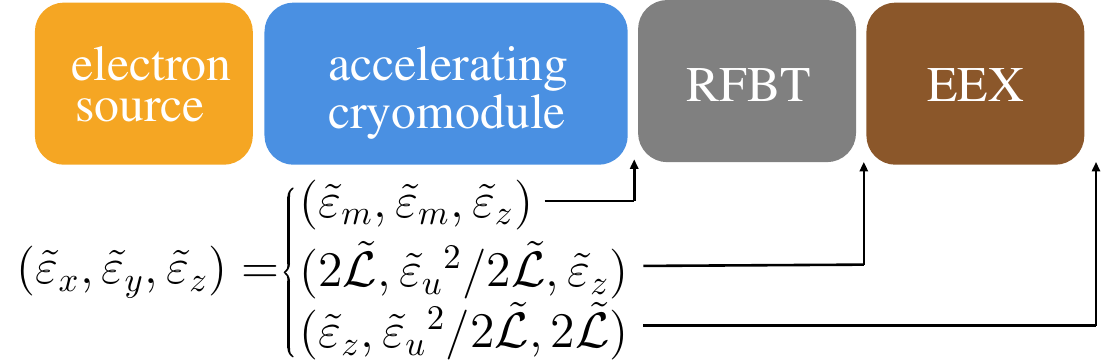}
\caption{\label{fig:block} Block diagram of the proposed damping-ring-free injector concept. The emittance partitions at the various stages along the injector are also listed. We defined $\rmsemit_{m}\equiv [\rmsemit_u^2+\tilde{\cal L}^2]^{1/2}$. See text for details.}
\end{figure}

A high-level block diagram of the proposed approach to realizing emittance partition consistent with LC requirements appears in Fig.~\ref{fig:block}. In a first stage, the electron beam is emitted from a cathode immersed in an axial magnetic field $B_{c}$ provided by a solenoidal field resulting in a ``magnetized" beam downstream of the magnetic-field region. The corresponding initial beam matrix $\Sigma\equiv \mean{{\boldsymbol{\mathcal Z}}{\boldsymbol{\mathcal Z}}^T}$ is~\cite{brinkmann-2001-a,kim-2003-a} 
\begin{eqnarray}\label{eq:sigma1}
\Sigma_i = R_{\mathrm{fr}} \Sigma_0 R_{\mathrm{fr}}^{T} =\left( \begin{array} { ccc} 
A &\widetilde{\cal L} J_2 & 0 \\ - \widetilde{\cal L} J_2 & A &  0\\ 0 & 0  & B \end{array} \right),
\end{eqnarray} 
where $\Sigma_0\equiv \mbox{diag}(\sigma_c^2, \rmsemit_c^2/\sigma_c^2, \sigma_c^2, \rmsemit_c^2/ \sigma_c^2,  \sigma_z^2, \rmsemit_z^2/ \sigma_z^2 )$ represent the uncorrelated beam matrix, and the matrix $R_{\mathrm{fr}}$ represents the fringe field experienced by the bunch as it exits the solenoidal field~\cite{brinkmann-2001-a}
\begin{eqnarray}
R_{\mathrm{fr}} = \left( \begin{array} { cccccc} 1 & 0 & 0 & 0 & 0 & 0\\ 0 & 1 & -\kappa_{0} & 0 & 0 & 0\\ 0 & 0  & 1 & 0 & 0 & 0 \\ \kappa_{0} & 0 & 0 & 1 & 0 & 0 \\  0 & 0 & 0 & 0 & 1 & 0 \\  0 & 0 & 0 & 0 & 0 & 1\end{array} \right), 
\end{eqnarray} 
where $\kappa_0\equiv \frac{eB_c}{2mc}$. 
In the r.h.s. of Eq.~\ref{eq:sigma1} the matrix $J_2\equiv \left( \begin{array} { cc } 0 & 1 \\ -1 & 0 \end{array} \right)$ is skew-symmetric simplectic matrix, $\widetilde{\cal L}\equiv \kappa_0 \sigma_c^2$, represents the beam magnetization (here $e$, $m$, and $c$ are respectively the electron charge, mass, and the velocity of light) which macroscopically characterizes the beam's average canonical angular momentum (CAM). Finally, the $2\times 2$ matrix $A$ is given by 
\begin{eqnarray}\label{eq:A}
A = \left( \begin{array} { cc} 
\sigma_c^2 & 0  \\ 
0  &   \frac{\rmsemit_c^2}{\sigma_c^2} + \kappa^2_0 \sigma_c^2
\end{array} \right),
\end{eqnarray} 
indicating that as the beam exits the magnetic-field region the conservation of CAM leads to a fully coupled beam with kinematical angular momentum $p_{\phi}=2mc{\cal L}$. It should also be noted that $[\mbox{det}(A)]^{1/2}=[\rmsemit_c^2+\tilde{\cal L}^2]^{1/2}$ represents the projected emittance in $(x,x')$ or $(y,y')$. 

Downstream of the electron source the beam is injected in a linac for acceleration. The acceleration is provided by cylindrical symmetric cavity which generally support a radially axisymmetric ponderomotive focusing~\cite{rosen-1994-a} thereby not affecting the form of the beam matrix described by  Eq.~\ref{eq:sigma1}. Downstream of the linac the beam is decoupled by applying a torque using three skew-quadrupole magnets~\cite{sun-2004-a} described by a total transfer matrix $M$. The final beam has an asymmetric transverse emittance partition~\cite{kim-2003-a} with corresponding beam matrix
\begin{eqnarray} \label{eq:Sigmaf}
\Sigma_{f} &=& M \Sigma_i { M^T} = \begin{pmatrix} 
\rmsemit_{x,f} T_x & 0 & 0 \\
0 & \rmsemit_{y,f} T_y & 0 \\
0 & 0 & \rmsemit_{z,f} T_z
\end{pmatrix},
\end{eqnarray}
where $T_{\ell} \equiv \left( \begin{array} { c c } \beta_{_{\ell}}  & -\alpha_{_{\ell}}  \\ - \alpha_{_{\ell}}  & \gamma_{_{\ell}}  \end{array} \right)$ with $\beta_{_{\ell}}>0$ being the betatron functions, $\alpha_{_{\ell}}\equiv -\frac{1}{2} \frac{d\beta_{\ell}}{ds}$ measures the phase-space linear correlation and $\gamma_{_{\ell}}\equiv (1+\alpha_{\ell}^2)/\beta_{\ell}$ so that its determinant is  $\mbox{det}(T_{\ell})=1$. The transverse flat-beam emittances are given by~\cite{burov-2002-a, kim-2003-a} 
\begin{eqnarray} \label{eq:eigen}
\rmsemit_{x,f} &\simeq & 2\widetilde{\cal L}\equiv \rmsemit_+, \mbox{~and}  \nonumber \\
\rmsemit_{y,f} &\simeq & \frac{\rmsemit_u^2}{2\widetilde{\cal L}}\equiv \rmsemit_-,  
\end{eqnarray}
where $\rmsemit_u\simeq [\rmsemit_c^2 + (\Delta\rmsemit)^2]^{1/2}$ should be understood as the uncorrelated emittance originating from the initial photocathode intrinsic emittance $\rmsemit_c$ but also accounting for other emittance-degrading effects (space charge effects, geometric nonlinearities and aberrations associated with the external focusing represented by the term $\Delta\rmsemit$) during acceleration and transport up to the entrance of the RFBT.

A proof of principle experiment demonstrated transverse emittance ratios $\varrho \simeq 100$~\cite{piot-2006-a} for a charge of 0.5~nC while a recent experiment has attained an emittance ratio of $\varrho \simeq 200$ for a 1-nC bunch~\cite{xu-2022-a}. \\

\begin{figure*}[t]
\includegraphics[width=2.0\columnwidth]{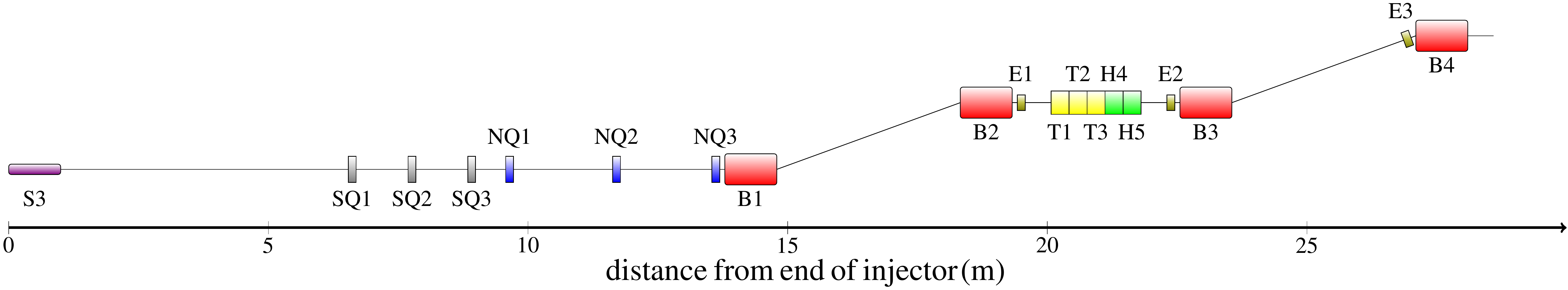}
\caption{\label{fig:fbteex_layout}  Overview of the emittance manipulation beamline combining the RFBT (skew-quadrupole magnets SQ1, SQ2, and SQ3) and EEX (from dipole magnet B1 to B4) insertions. The label ``SQ$i$" and ``NQ$i$" refer to skew- and normal-quadrupole magnets, ``B$i$" and ``E$i$" are dipole nd sextupole magnets. The elements "T$i$" and ``H$i$" respectively refer to transverse-deflecting and accelerating SRF cavities; ``S3" is a solenoidal magnetic lens.}
\end{figure*}

The second stage of the proposed photoinjector consists of exchanging the horizontal and longitudinal phase spaces using a EEX beamline. The design of such beamline was extensively discussed in, e.g., Refs~\cite{cornacchia-2002-a,emma-2006-a,nanni2015}. A solution for such a EEX beamline consists of a deflecting cavity flanked by two dispersive sections. In order to ensure the  transfer matrix in  is 2x2-block anti-diagonal in $(x,x',z,\delta)$, the deflecting voltage $V_{\perp}$ is related to the dispersion $\eta$ generated by the upstream dispersive section following $1+\kappa \eta=1$, where $\kappa\equiv \frac{keV_{\perp}}{E_b}$ is the deflecting strength and $k\equiv 2\pi/\lambda$ (with $\lambda$ being the deflecting-mode wavelength). Under such a condition the general transfer matrix of an EEX beamline is 
\begin{eqnarray} \label{eq:EEX}
R_{EEX} &=& \begin{pmatrix} 
0 & 0 & F \\
0 & E & 0 \\
F^{-1} & 0 & 0
\end{pmatrix}. 
\end{eqnarray}
A simple implementation of an EEX beamline consists of deflecting cavity flaned by two identical dispersive section arranged as dogleg~\cite{emma-2006-a}. In such a case the matrix $F$ is 
\begin{eqnarray}
F &=& \begin{pmatrix} 
-\frac{L}{\eta} & \eta-\frac{\xi L}{\eta}   \\
-\frac{1}{\eta}  & -\frac{\xi}{\eta}   \\
\end{pmatrix},
\end{eqnarray}
where $\eta$ and $\xi$ are respectively the horizontal and longitudinal dispersion downstream of one dogleg and $L$ its length. Such EEX beamlines have demonstrated near-ideal emittance exchange~\cite{ruan-2011-a} and the formation of temporally-shaped beams~\cite{sun-2010-a,ha-2017-a}.

The final beam matrix downstream of the EEX is 
\begin{eqnarray}
\Sigma_e &=& M \Sigma { M^T} = \begin{pmatrix} 
\rmsemit_{z,f} T'_x & 0 & 0 \\
0 & \rmsemit_{y,f} T'_y & 0 \\
0 & 0 & \rmsemit_{x,f} T'_z
\end{pmatrix},
\end{eqnarray}
where $T'_{\ell}$ (with $\ell=x, y, z$) assumes the same form as the matrix $T_{\ell}$ introduced in Eq.~\ref{eq:Sigmaf}. Consequently, the final normalized-emittance partition is  
\begin{eqnarray}
(\varepsilon_{x,e}, \varepsilon_{y,e}, \varepsilon_{z,e})=(\varepsilon_{z,0}, \frac{\varepsilon_{u}^2}{2{\cal L}}, 2{\cal L}), 
\end{eqnarray}
where ${\cal L}\equiv \gamma\widetilde{\cal L}$ following our earlier convention for emittance.

\subsection{Deviations from linear transformation}\label{sec:deviation}
The process described in the previous Section~\ref{ssec:linear} idealizes the emittance partitioning and exchange by describing the associated transform with linear transfer matrices and ignoring collective effects. In this section, we briefly review some limitations of the process and corrections that were considered for the design simulated in Section~\ref{sec:simulation} and diagrammed in Fig.~\ref{fig:fbteex_layout}. 
First, it should be noted that in our configuration we constrain the beam to have a low fractional energy spread before the RFBT which results in insignificant chromatic aberration and near-ideal transfer of eigenemittance to transverse emittance.\\

As far as the EEX is concerned one critical deviation from the matrix model discussed in the previous section comes from the thick-lens matrix of the deflecting cavity (labeled as T1-3 in Fig.~\ref{fig:fbteex_layout}) which introduces a coupling element between the horizontal and longitudinal DOF~\cite{cornacchia-2002-a} and breaks the block anti-diagonal form of $R_{EEX}$ given by Eq.~\ref{eq:EEX}. However, the cancellation of this term was shown to be possible using an accelerating cavity operating at zero crossing~\cite{zholents-2011-a,xiang-2011-a}. Consequently, accelerating cavities were introduced (H4-5 in Fig.~\ref{fig:fbteex_layout}) downstream of the deflecting cavities. 

The beam dynamics in the EEX section is impacted by second-order effect. In Ref.~\cite{emma-2006-a} it was pointed out that a proper LPS chirp could mitigate second-order aberration. In our setup given the targeted vertical emittance the introduction of the chirp would have to be done with another linac module located between the RFBT and EEX as a chirp at the entrance of the RFBT would impact the small vertical emittance due to chromatic aberration in the RFBT.Given the need to minimize the final horizontal emittance, we  follow the analysis detailed in Ref.~\cite{ha-2016} to understand the source of possible final horizontal-phase-space diluation. We start by considering the phase-space coordinate of an electron downstream of the first dogleg (consisting of dipole magnets B1 and B2) we have 
\begin{eqnarray} 
x _ { 1 } &=& x _ { 0 } + L x _ { 0 } ^ { \prime } + \eta \delta _ { 0 } + T _ { 122 } x _ { 0 } ^ { \prime 2 } + T _ { 126 } x _ { 0 } ^ { \prime } \delta _ { 0 } \nonumber \\
&&  +  T _ { 133 } y _ { 0 } ^ { 2 } +  T _ { 134 } y _ { 0 } y _ { 0 } ^ { 2 } + T _ { 144 } y _ { 0 } ^ { \prime 2 } + T _ { 166 } \delta _ { 0 } ^ { 2 } \\ 
x _ { 1 } ^ { \prime } &=& x _ { 0 } ^ { \prime } + T _ { 233 } y _ { 0 } ^ { 2 } + T _ { 234 } y _ { 0 } y _ { 0 } ^ { \prime } + T _ { 244 } y _ { 0 } ^ { 2 }
\end{eqnarray}
for the horizontal phase space. The longitudinal phase-space coordinates are
\begin{eqnarray} 
z _ { 1 } &=&  \eta x _ { 0 } ^ { \prime } + z _ { 0 } + \xi \delta _ { 0 } + T _ { 522 } x _ { 0 } ^ { \prime 2 } + T _ { 526 } x _ { 0 } ^ { \prime } \delta _ { 0 } \nonumber \\
&& + T _ { 533 } y _ { 0 } ^ { 2 } + T _ { 534 } y _ { 0 } y _ { 0 } ^ { \prime } + T _ { 544 } y _ { 0 } ^ { \prime 2 } + T _ { 566 } \delta _ { 0 } ^ { 2 } \\ 
\delta _ { 1 } &=& \delta _ {0}. 
\end{eqnarray}
In the latter equations, the subscript $_0$ indicates the coordinate upstream of B1, and the $T_{ijk}$ are the usual second-order aberration coefficients~\cite{brown-1975-a} associated with one dogleg~\footnote{The nonlinear aberrations arising from the deflecting cavity are ignored in this section for sake of simplicity. Their inclusions do not affect the discussion and overall aberration-correction method.}. 

Finally, the horizontal coordinates after the EEX section are given by
\begin{eqnarray} \label{eq:x2final}
x_{2} &=& x _ { 0 }  + T _ { 166 } \delta_{2} ^ { 2 } + L x _ { 0 } ^ { \prime } + L _ { d } x _ { 0 } ^ { \prime } + \delta _ { 0 } \eta + L  x_{2}^{\prime}  + L _ { d }  x_{2}^{\prime} \nonumber \\  
&& + T _ { 122 } x _ { 0 } ^ { \prime } { } ^ { 2 } + T _ { 122 }  x_{2}^{\prime} {} ^ { 2 } + \eta \delta_{2} + x _ { 0 } ^ { \prime } \left( L _ { a } + L _ { c } \right) \nonumber \\
&& +  \kappa \left( L _ { a } + \frac { L _ { c } } { 2 } \right) \left( T _ { 522 } x _ { 0 } ^ { \prime } { } ^ { 2 } + \eta x _ { 0 } ^ { \prime } + z _ { 0 } + \delta _ { 0 } \xi \right) \nonumber \\
&& + T _ { 126 }  x_{2}^{\prime} \delta_{2} \label{eq:final_x}\\
x_{2}^{\prime} &=& x _ { 0 } ^ { \prime } + \kappa \left( T _ { 522 } x _ { 0 } ^ { \prime } { } ^ { 2 } + \eta x _ { 0 } ^ { \prime } + z _ { 0 } + \delta _ { 0 } \xi \right)
\end{eqnarray}
where $\delta_{2} \equiv  \delta _ { 0 } + \kappa \left( x _ { 0 } + L x _ { 0 } ^ { \prime } + L _ { d } x _ { 0 } ^ { \prime }  + \delta _ { 0 } \eta + T _ { 122 } x _ { 0 } ^ { \prime } { } ^ { 2 } \right) + \frac { L _ { c } \kappa x _ { 0 } ^ { \prime } } { 2 }$, and $   x_{2}^{\prime} = x _ { 0 } ^ { \prime } + \kappa \left( T _ { 522 } x _ { 0 } ^ { \prime } { } ^ { 2 } + \eta x _ { 0 } ^ { \prime } + z _ { 0 } + \delta _ { 0 } \xi \right)$ are the $\delta$ and $x^{\prime}$ coordinates after second dogleg.
In latter equation we neglected geometric aberrations arising from the coupling with the $(y,y')$ given the very low vertical emittance.  Likewise, we ignore the  $T_{126}$ and $T_{526}$ terms associated with the first dogleg since the initial $x_0^{\prime}-\delta_0$ correlation is small (ideally vanishing).

The $T _ { 122 } x _ { 0 } ^ { \prime } { } ^ { 2 }$ and $T _ { 522 } x _ { 0 } ^ { \prime } { } ^ { 2 }$ terms in the final horizontal coordinates can be minimized by imposing a large $\beta_{x}$ at the entrance of EEX. The rest of the second order terms related to $\delta_{2}$ and $x_{2}^{\prime}$ can be reduced with a initial correlation in $(x_{0},x _ { 0 } ^ { \prime })$ and $(z_{0},\delta_{0})$ to produce a horizontal and longitudinal beam waist at the center of the TDC so the quantity $T_{166}\delta_{2}^{2}$,  $T_{122}x_{2}^{\prime}{}^{2}$ and $T_{126}x_{2}^{\prime}\delta_{2}$ in Eq.~\ref{eq:x2final} are minimized. Finally, the $(x_{2},x_{2}^{\prime})$ coordinate downstream of B4 can be written as\begin{eqnarray}
x_{2} &=& \eta\delta_0 - \frac{L+L_{d}+L_{a}+L_{c}}{2\eta }(z_{0}+\xi\delta_{0})\\ 
x_{2}^{\prime} &=& \frac{-1}{\eta}(z_{0}+\xi\delta_{0}). 
\end{eqnarray}
The previous equation is obtained by enforcing the condition $1+\eta\kappa=0$ required for emittance exchange.

\section{Numerical proof of concept for producing beam with ILC-like parameters\label{sec:simulation}}

In this section we apply the concept devised in the previous section to the case of the ILC to produce an emittance partition similar to the one produced downstream of the ILC damping ring~\cite{phinney-2007-a}; see Table~\ref{tab:lcparam}. The design philosophy focuses on designing an injector capable of minimizing the beam emittance along all d.o.f's upstream of the RFBT, and then optimizing the emittance repartitioning in the RFBT and emittance-exchange process in the EEX beamlines. Each of these steps is discussed below. 

\subsection{Beam generation~\label{sec:injetor}}
\begin{figure}[hh!!!!!]
\centering
\includegraphics[width=1.0\columnwidth]{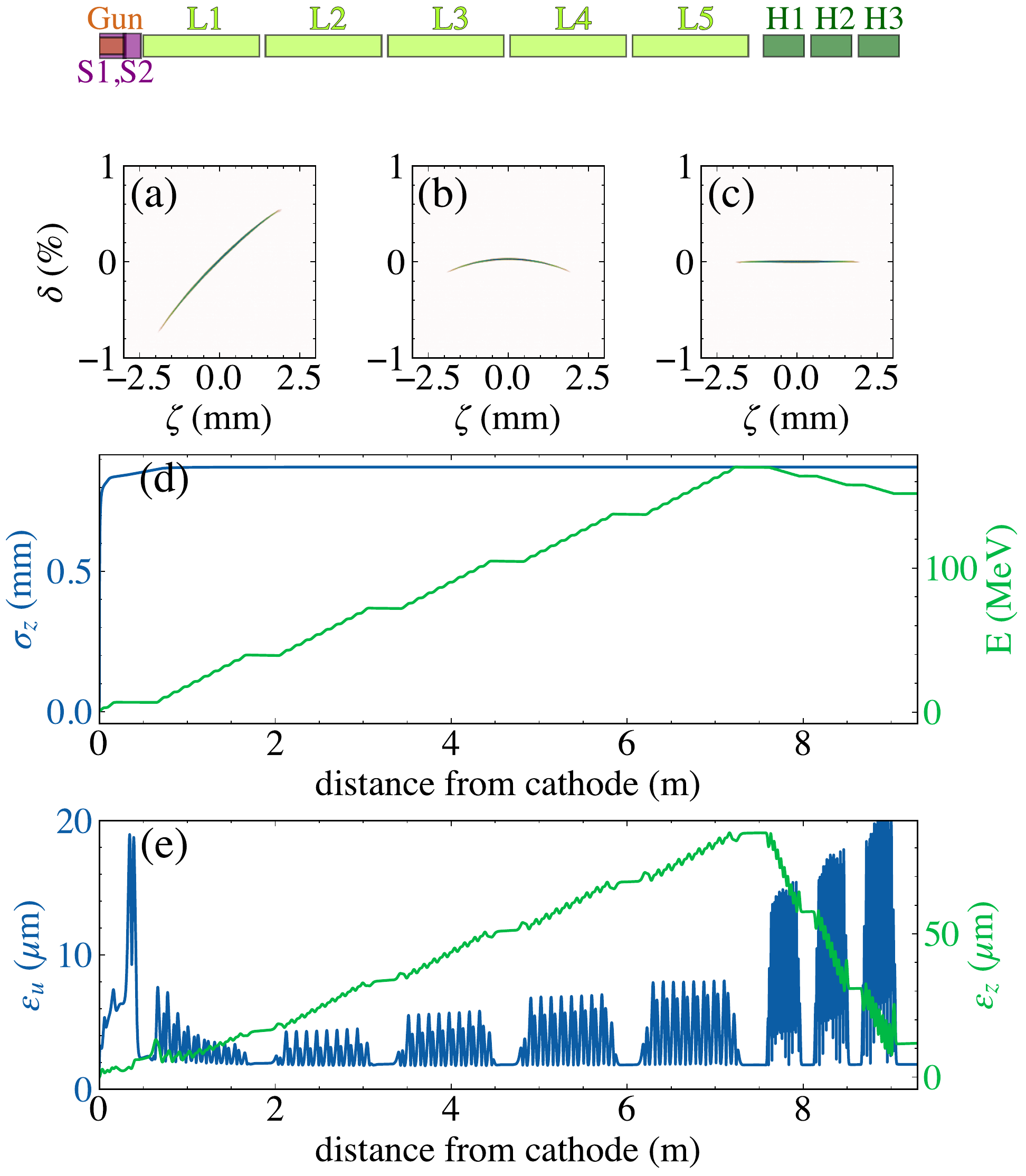}
\caption{\label{fig:injector_evo}  Photoinjector diagram (upper schematics) and snapshots of the LPS distribution at $z=1.88$ (a), $7.48$ (b), and $9.3$~m (c) from the photocathode. Evolution of the beam energy and RMS bunch length (d) and corresponding 4D transverse and longitudinal emittances (e). In the upper block diagram, S1 and S2 respectively refer to the solenoidal magnetic lenses, L1-5 are the 1.3-GHz SRF cavities, and H1-3 represent the 3.9-GHz SRF cavities. In plots (a-c) and throughout this paper, $\zeta>0$ corresponds to the head of the bunch.}
\end{figure}

The conceptual design of the photoinjector beamline from the photocathode surface up to the entrance of the RFBT is diagrammed in Figure~\ref{fig:injector_evo}. The injector beamlines was modeled using the particle-in-cell beam-dynamics program {\sc impact-t}~\cite{qiang-2006-a}. The electron source consists of a $1+\frac{1}{2}$-cell RF gun operating at $f_0=1.3$~GHz operating with a peak field on the cathode of $E_c=60$~MV/m. The downstream linac consists of five TESLA-type 9-cell superconducting RF (SRF) cavities operating at a peak field of $E_L=60$~MV/m (corresponding to an accelerating gradient $G_L\simeq E_L/2 \simeq 30$~MV/m consistent with ILC demonstrated requirement of $G_L = 31.5$~MV/m~\cite{broemmelsiek-2018-a}). The RF gun is nested in a pair of solenoidal lenses to control the beam emittance. The beamline parameters [laser spot radius, solenoid (S1 and S2) strengths and locations, field amplitude and phase of L1] were optimized to minimize the transverse uncorrelated emittance $\varepsilon_u$ and maximize the eigenemittance ratio $\varrho\equiv \varepsilon_+/\varepsilon_-$ at the exit of the L1. To ensure a minimal longitudinal emittance and space-charge effects, we considered a spatiotemporally shaped laser pulse with uniform three-dimensional ellipsoidal intensity  distribution~\cite{li-2008-a,li-2009-a}. 

The photoemitted electron beam mirrors the laser distribution thereby prducing space-charge fields with a linear dependence on the spatial coordinate within the ellipsoidal bunch~\cite{lapostolle-1965-a,luiten-2004-a}. The linear space-charge force mitigates emittance dilution and imparts a significant chirp in the longitudinal phase space (LPS). Additionally, the resulting bunch length [$\sigma_z\simeq 0.87$~mm; see Fig.~\ref{fig:injector_evo}(a)] leads the LPS to develop a quadratic correlation induced by the RF waveform; see Fig.~\ref{fig:injector_evo}(b). The linac cavities (L2-5) are operated $\varphi_L = 2^{\circ}$ off-crest to remove the linear LPS correlation after acceleration to 151~MeV; see Fig.~\ref{fig:injector_evo}(b). The 1.3-GHz linacs are followed by a 3rd-harmonic accelerating-cavity module operating at $f_H=3f_0=3.9$~GHz to correct the quadratic correlation in the LPS and reduce the longitudinal emittance. The module comprises three SRF 3rd-harmonic cavities (H1-3) with a similar design as discussed in Ref.~\cite{bertucci-2019-a}. The cancellation of the quadratic correlation gives an 8 fold decrease in the longitudinal emittance to a final value of $\varepsilon_z\simeq 11.78$~\textmu{m}; see Fig.~\ref{fig:injector_evo}(e). The beamline parameters and resulting beam-emittance partitions are summarized in Table~\ref{tab:ilcmain}. 

%
\begin{table}[hhhh!!]
\caption{Beamline settings for the proposed photoinjector and achieved normalized-emittance values at the end of the beamline. The quantities $\varepsilon_{\pm}\equiv \gamma \rmsemit_{\pm}$ where $\rmsemit_{\pm}$ is defined in Eq.~\ref{eq:eigen} \label{tab:ilcmain}}
\begin{center}
\begin{tabular}{l c c c}\hline\hline
parameter  & symbol & value & unit \\
\hline
charge  & Q &  3.2  & nC     \\
laser pulse full (and rms) duration   & $\tau_l$ &  10 (2.24)  & ps     \\
laser rms  spot size & $\sigma_c$ &  1.93  & mm     \\
thermal emittance & $\varepsilon_{c}$ &  1.634  & $\mu$m \\
magnetic field on cathode    & $B_c$  & 226 & mT  \\
laser/gun launch phase  & $\varphi_0$\footnote{emission phase wrt to zero-crossing.}  &  50 & deg \\
peak E field on cathode  & $E_0$  &  60 &  MV/m  \\
L2-L5 off-crest phase  & $\varphi_L$  &  2   &   deg \\
linac peak electric field  & $E_L$ &  60 &   MV/m \\
H1-H3 off-crest phase  & $\varphi_H$  &  178.68   &   deg \\
H1-H3 peak electric field  & $E_H$ &  34   &   MV/m \\
\hline
total beam energy & $E_b $ &  151   &     MeV \\
longitudinal emittance  & $\varepsilon_z$  &  11.78   &     \textmu{m} \\
transverse eigenemittance  & $\varepsilon_-$  &  6.84   &  \text{nm} \\
transverse eigenemittance  & $\varepsilon_+$  &  493.4   &   \textmu{m} \\
transverse uncorrelated emittance & $\varepsilon_u$  & 1.85  &   \textmu{m} \\
magnetization  & ${\cal L}$  & 246.7  &   \textmu{m} \\
\hline
\hline
\end{tabular}
\end{center}
\end{table}


\subsection{Emittance Manipulation}

The emittance-manipulation beamline comprising the RFBT and EEX sections was simulated using {\sc elegant}~\cite{borland-2000-a}. The simulations account for higher-order aberrations and bunch self-interaction due to coherent synchrotron radiation (CSR). The beamline is located just after the photoinjector displayed in Fig.~\ref{fig:injector_evo},  at an energy of $\sim 151$~MeV. Downstream of the injector, the magnetized beam is focused by a solenoid into RFBT sections where three skew quadrupoles remove the angular momentum of the magnetized beam and transform the magnetized beam into flat beams with emittance partition downstream of the RFBT 
\begin{eqnarray}
(\varepsilon_{x,f}, \varepsilon_{y,f}, \varepsilon_{z,f})=(493.40, 7.17\times 10^{-3}, 11.82)~\mbox{\textmu{m}}. \nonumber 
\end{eqnarray}
This emittance partition confirms that the mapping of the transverse eigenemittances listed in Table~\ref{tab:ilcmain} to transverse emittance is near ideal (the emittance dilution associated with the mapping $\varepsilon_- \xrightarrow{} \varepsilon_y$  is 4.8\%) and the longitudinal emittance is preserved (relative emittance growth of 0.3\%).  
The flat beam is then matched into the EEX beamlines with NQ1-3 to meet the Courant-Snyder parameters requirement described in Section~\ref{sec:deviation}. The condition for the $(z_{0},\delta_{0})$ correlation is not imposed as we found the contribution of the $T _ { 122 } x _ { 2 } ^ { \prime } { } ^ { 2 }$ term in Eq.~\ref{eq:final_x} is insignificant for our beam parameters. The EEX beamline consists of two doglegs each with dipole bending angles of $(+2^{\circ},-2^{\circ})$, three 3.9-GHz deflecting cavities, and two 3.9-GHz accelerating cavities. The use of multiple SRF cavities is required given the demonstrated cavity performance (maximum achievable deflecting or accelerating voltage) and our requirements. Aside from canceling the thick lens effect of TDC, the accelerating cavities are also used to partially compensate for the correlated energy spread induced by CSR. Additionally, three sextupole magnets (labeled as E1-3) are inserted in the EEX beamline to correct the nonlinearities arising from the deflecting and accelerating 3.9-GHz cavities. The voltages of the TDC and third harmonic cavities, along with the strengths of the sextupole magnet, were numerically optimized to minimize the final horizontal emittance downstream of the EEX beamline. The optimized settings for cavities and magnets appear in Table~\ref{tab:ilcele}. 

\begin{figure}[hh!!!!!]
\centering
\includegraphics[width=1.0\columnwidth]{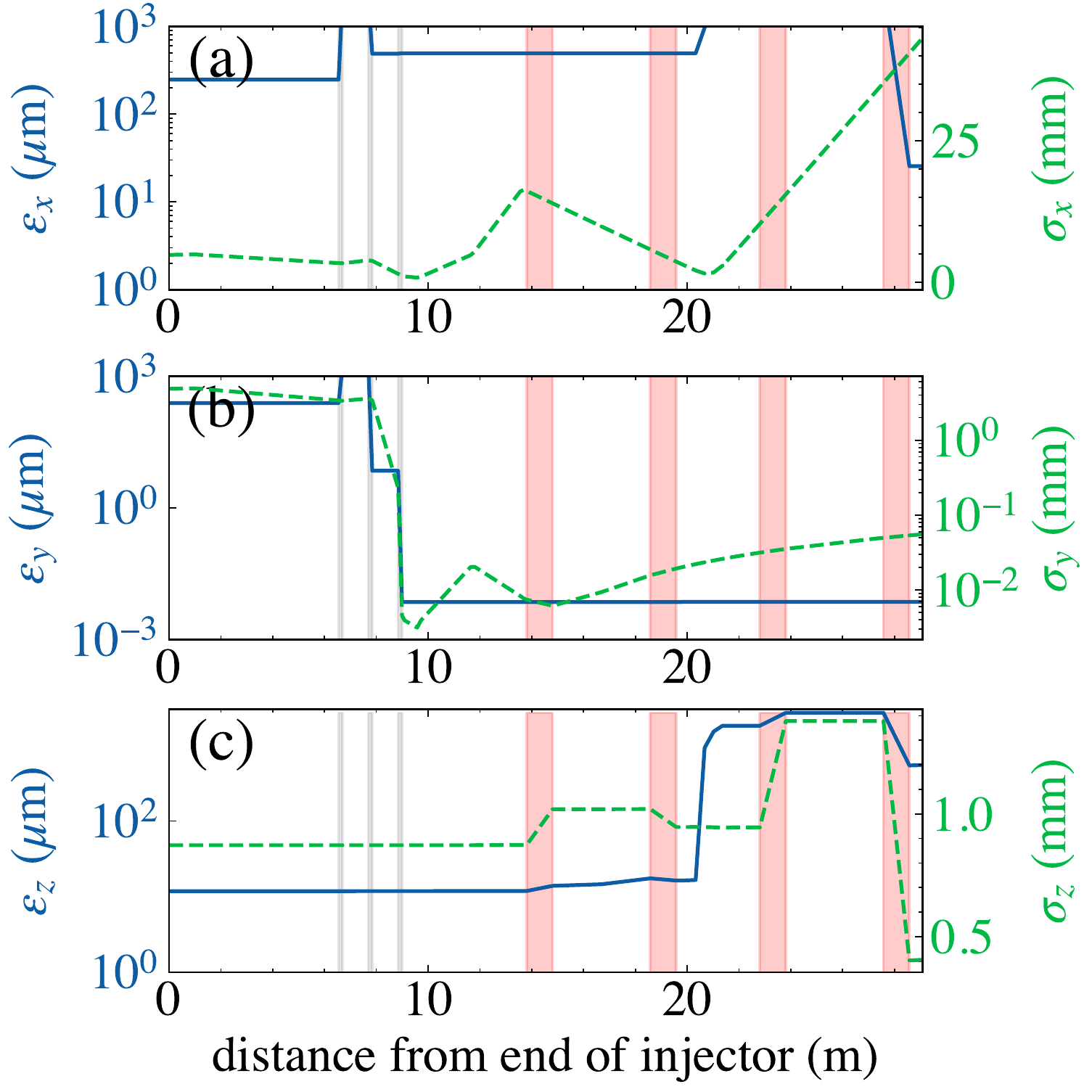}
\caption{\label{fig:eex_emit_evo}  Evolution of the horizontal (a), vertical (b) and longitudinal (c) emittance (blue traces) and bunch size (green dashed traces) along the emittance manipulation beamline (combining the RFBT and EEX transformations). The vertical shaded bands indicate the locations for the RFBT's skew quadrupoles (grey lines at distances $<10$~m are for SQ1-3) and dipole magnets (red bands from $\sim14$~m to the end of the beamline are for B1-4) associated with the EEX beamline; see Fig.~\ref{fig:fbteex_layout}.}
\end{figure}

\begin{figure}[hh!!!!!]
\centering
\includegraphics[width=1.0\columnwidth]{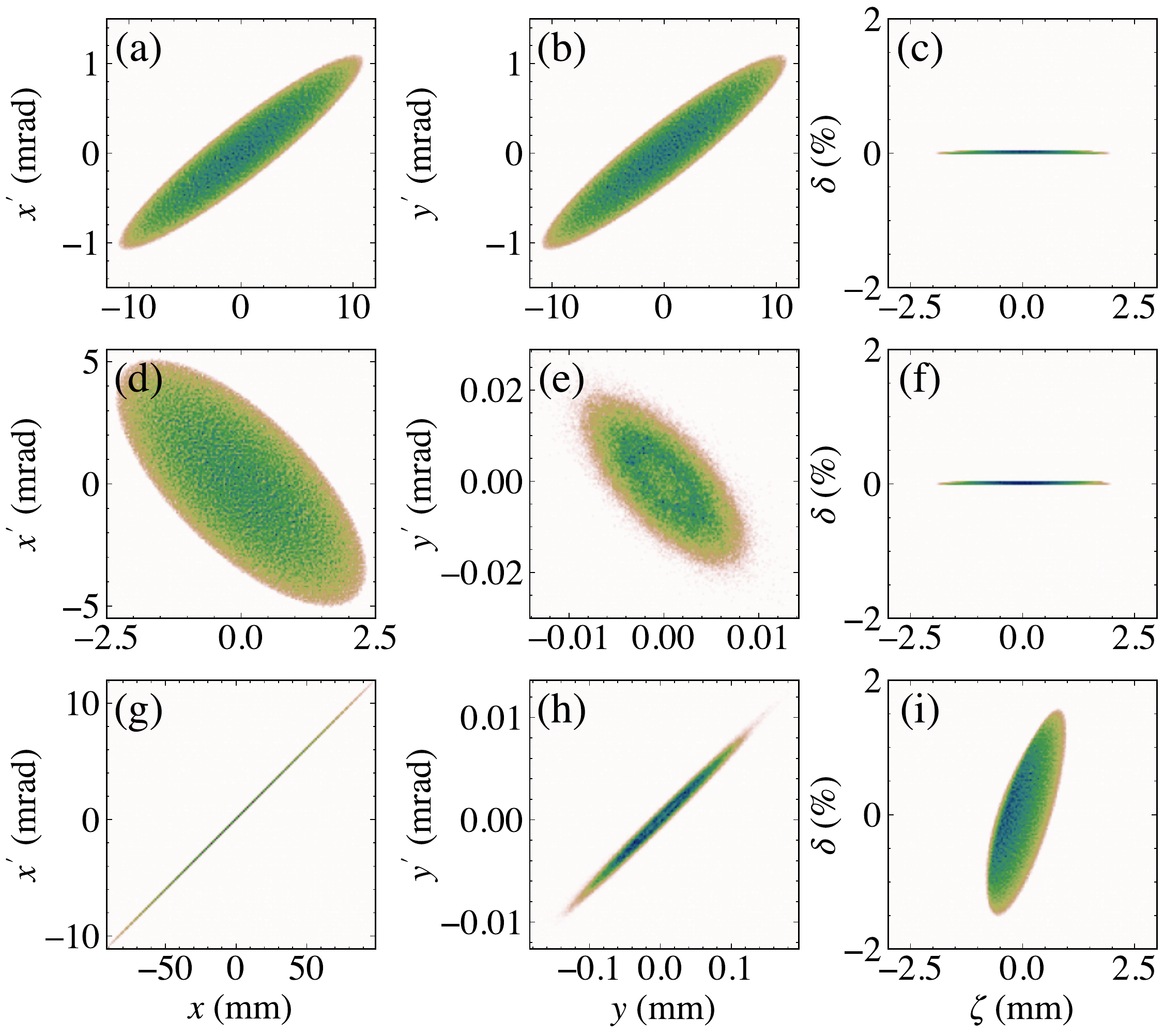}
\caption{\label{fig:all_ps} Horizontal (a,d,g), vertical (b,e,h) and longitudinal (c,f,i) phase space upstream of the RFBT (a,b,c), upstream of the EEX (d,e,f) and at the exit of the EEX (g,h,i).}
\end{figure}

The evolution of the beam emittances along the emittance-manipulation section is presented in Fig.~\ref{fig:eex_emit_evo} and confirms a final emittance partition of 
\begin{eqnarray}
(\varepsilon_{x,e}, \varepsilon_{y,e}, \varepsilon_{z,e})=(25.47, 7.26\times 10^{-3}, 546.34)~\mbox{\textmu{m}} \nonumber 
\end{eqnarray}
was attained corresponding to a 6D brightness ${\cal B}_6\simeq 31.7 $~pC/(\textmu{m}$^3$). This 6D brightness is a factor of $\sim 3$ higher than the one listed under ``RF gun" in Table~\ref{tab:lcparam} most likely due to the use of a 3D ellipsoidal photocathode-laser distribution in the present work while Ref.~\cite{krasilnikov-2012-a} employs a uniform-cylinder laser distribution. Snapshots of the phase-space distributions at different stages of the beam generation and manipulation along the beamline appear in Fig.~\ref{fig:all_ps}. \\

\begin{figure}[hh!!!!!]
\centering
\includegraphics[width=1.0\columnwidth]{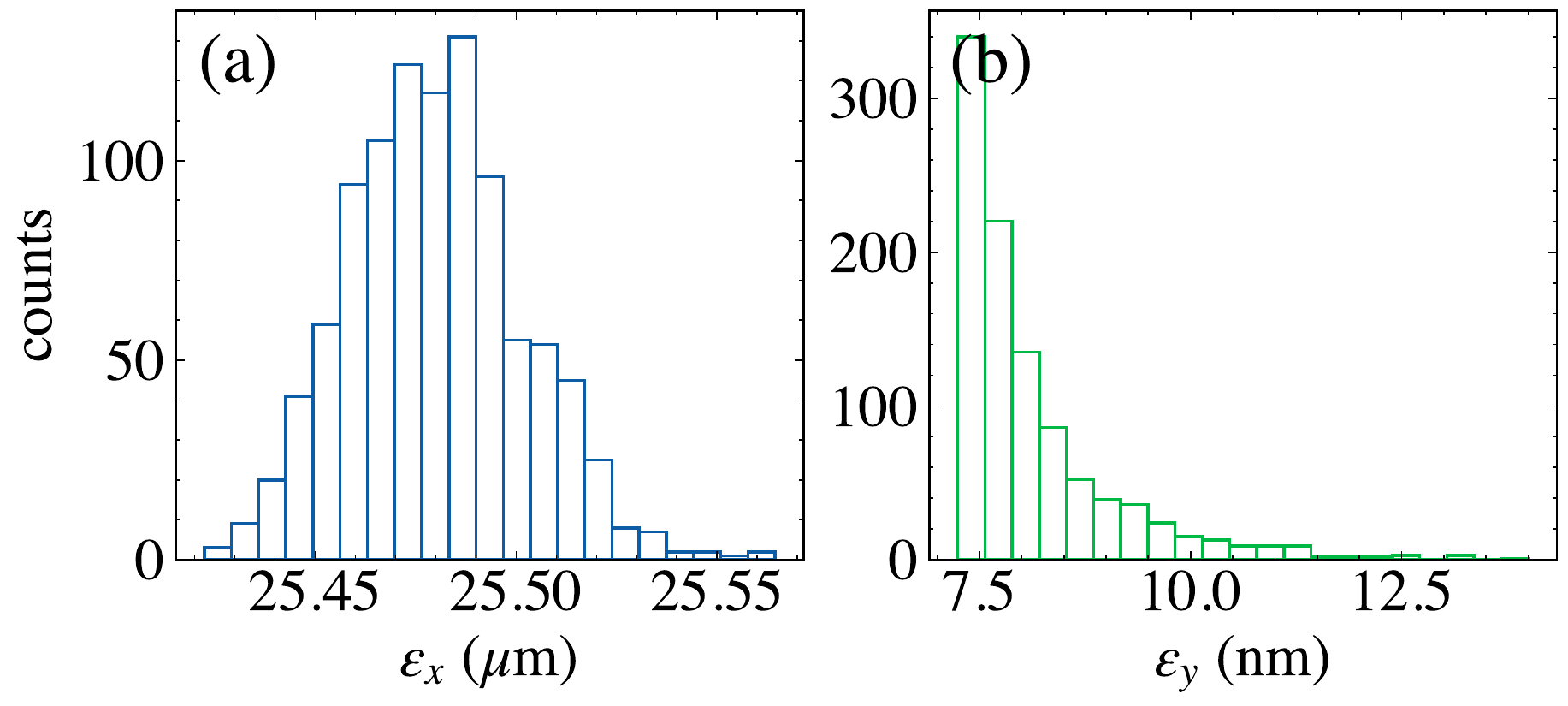}
\caption{\label{fig:jitter_emit}  Histogram of final horizontal (a) and vertical (b) emittances simulated downstream of the EEX beamline for 1000 realizations of SRF-cavity random phase and amplitude jitters.}
\end{figure}

We evaluated the robustness of the proposed design and the sensitivity of the final transverse emittances to shot-to-shot jitters associated with amplitude and phase stability of the SRF cavities via start-to-end simulations. Specifically, we performed 1000 start-to-end simulations with different random realizations of the RF amplitude and phase for all the SRF cavities. The amplitude and phase values were randomly generated with a normal distribution with respective rms jitter of 0.01\% (fractional deviation from nominal-amplitude settings) and 0.01 degree (for the 1.3~GHz cavities) and 0.03 deg (for the 3.9~GHz cavities). These tolerances are consistent with the performances of the low-level RF system at the European X-ray FEL~\cite{omet-2018-a}. These jitter studies confirm that the associated transverse-emittance fluctuations are acceptable -- i.e.  $\varepsilon_x=25.48 \pm 0.02$~\textmu{m} and $\varepsilon_y=8.13 \pm 0.98$~nm; see corresponding histogram in Fig.~\ref{fig:jitter_emit}.

\begin{table}[hhhh!!]
\caption{Operating parameters RFBT and EEX beamline, the magnet names refer to Fig.~\ref{fig:fbteex_layout}.\label{tab:ilcele} }
\begin{center}
\begin{tabular}{l l l}\hline\hline\
parameter  & value & unit \\
\hline
skew quadrupole magnet SQ1 &  $k_1=3.71$   & m$^{-1}$     \\
skew quadrupole magnet SQ2  &  $k_1=-7.08$  & m$^{-1}$     \\
skew quadrupole magnet SQ3  &  $k_1=15.76$  & m$^{-1}$     \\
sextupole magnet E1 &  $k_2=-15.67$  & m$^{-2}$     \\
sextupole magnet E2 &  $k_2=-1.08$   & m$^{-2}$     \\
sextupole magnet E3 &  $k_2=-0.03$   & m$^{-2}$     \\
doglegs dispersion $\eta$ &  -1.67  & m     \\
TDC section kick strength $\kappa$ &  6  & m$^{-1}$ \\
dipole magnet B1-B4 angles  &  2  & deg     \\
T1 deflecting voltage  &  3.72  & MV     \\
T2 deflecting voltage  &  3.72  & MV     \\
T3 deflecting voltage  &  3.66  & MV     \\
H4 accelerating voltage  &  5.81  & MV     \\
H5 accelerating voltage  &  5.91  & MV     \\
\hline
\hline
\end{tabular}
\end{center}
\end{table}

\subsection{Spin dynamics}

The present requirements from high-energy physics call for 80\% spin-polarized electron beams. The $e^-$/$e^+$ bunch charge ranges from fC to nC depending on the LC technology choice~\cite{alegro-2019-a}. In most of the designs, the polarized electron beam is produced via photoemission from semiconductor Gallium-Arsenide (GaAs) photocathodes placed in a DC-gun~\cite{sinclair-2007-a}. Operation of a Gallium-Arsenide (GaAs) photocathodes in an RF gun remains a challenge and has been the subject of intense research~\cite{aulenbacher-1996-a,aleksandrov-1998-a,fliller-2005-a}. 
The photoinjector is expected to produce a longitudinally spin-polarized electron beam with most of the electrons' spin vector $\mathbf{S}=S_z\hat{\mathbf{z}}$.

The evolution of the spin in an externally-applied magnetic field $ \mathbf { B }$ can be described by the classical spin vector $\mathbf{S}$ under the action of a semiclassical spin precession vector $\mathbf{\Omega}$ via the BMT equation~\cite{bargmann-1959-a}
\begin{eqnarray}
\frac { d \mathbf{ S } } { d t } = \mathbf{ S } \times \mathbf{ \Omega }
\end{eqnarray}
with, 
\begin{eqnarray}
\boldsymbol{ \Omega } &=& \frac { e } { m } \left[ \left( a + \frac { 1 } { \gamma } \right) \mathbf { B } - \frac { a \gamma } { \gamma + 1 } ( \boldsymbol{\beta} \cdot \mathbf { B } ) \boldsymbol { \beta } \right. \nonumber \\
&& \left. - \left( a + \frac { 1 } { \gamma + 1 } \right) \boldsymbol{ \beta } \times \frac { \mathbf{ E } } { c } \right],
\end{eqnarray}
where $a$ is anomalous magnetic moment and $\boldsymbol{\beta}\equiv \frac{\mathbf{v}}{c}$ with $\mathbf{v}$ being the velocity.
\begin{figure}[t!!!!!]
\centering
\includegraphics[width=0.90\columnwidth]{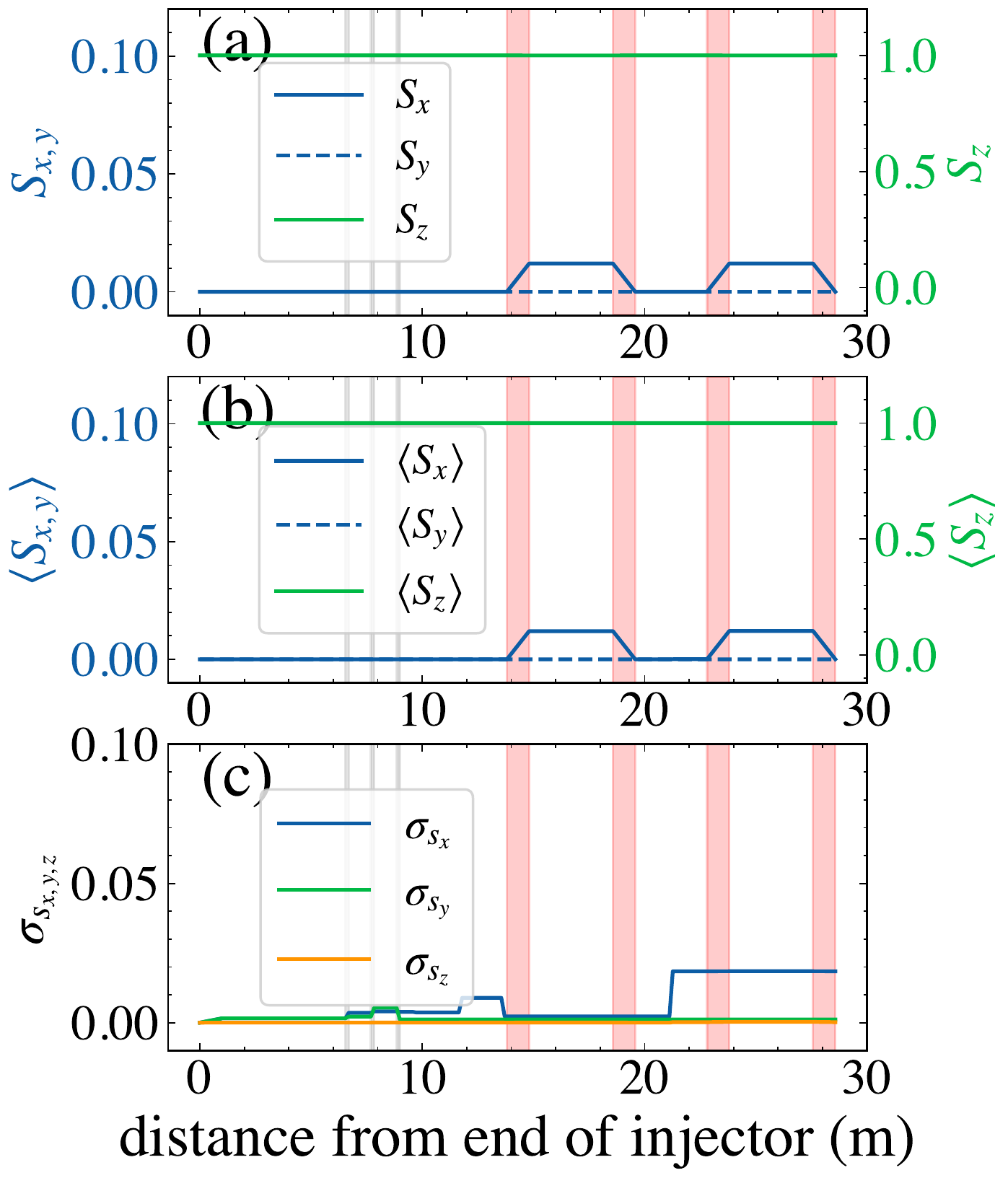}
\caption{\label{fig:spin_evo}  Evolution of the spin components along the emittance-manipulation beamline. Spin components associated with the reference particle $\mathbf{S}^T=(S_x, S_y, Sz)$ (a), statistical average $\mean{\mathbf{S}}$ (b) and RMS value $\mean{\mathbf{S}^2}^{1/2}$  (c) computed over the macroparticle distribution.   The vertical shaded bands indicate the locations for the RFBT's skew quadrupoles (grey lines at distances $<10$~m are for SQ1-3) and dipole magnets (red bands from $\sim14$~m to the end of the beamline are for B1-4) associated with the EEX beamline; see Fig.~\ref{fig:fbteex_layout}.}
\end{figure}

The spin dynamics of the particle distribution was investigated with the beam-dynamics program {\sc bmad}~\cite{bmad2006} which implements a Romberg integration of the spin rotation matrix. Figure~\ref{fig:spin_evo} presents the evolution of spin-vector components through the RFBT and EEX sections shown in Fig.~\ref{fig:fbteex_layout}. The initial conditions are such that the beam is 100\% longitudinally spin-polarized ${\pmb S}^T=(0,0,1)$. The simulation indicate that the RFBT does not impact the spin (no depolarization is observed) while the EEX beamline yield a small depolarization with final mean and RMS longitudinal spin values being respectively  $\mean{{\pmb S}_e}^T=(5.41\times10^{-5}, -1.39\times10^{-8}, 0.99)$ and $(\sigma_{S_{x,e}}, \sigma_{S_{y,en}}, \sigma_{S_{z,e}})=(1.84\times10^{-2}, 1.12\times10^{-3}, 1.81\times10^{-4})$. confirming that the longitudinal depolarization $\frac{\sigma_{S_{z,e}}}{\mean{S_{z,e}}}\sim {\cal O}( 10^{-4})$ is insignificant .

\subsection{Enhanced Luminosity}
\begin{figure}[hh!!!!!]
\centering
\includegraphics[width=1.0\columnwidth]{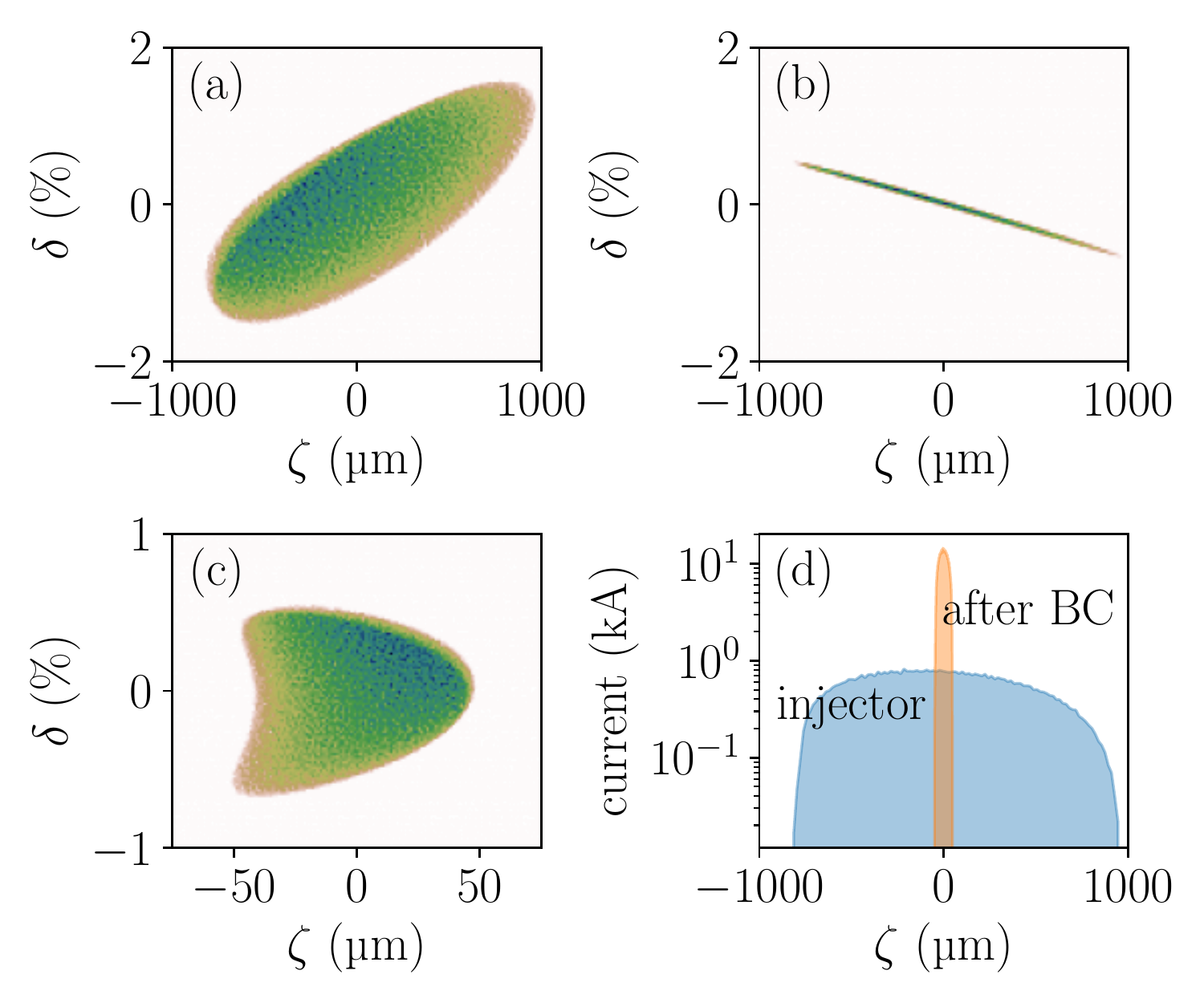}
\caption{\label{fig:LPScompression}  Snapshots of the LPS distributions at the exit of the photoinjector (a), after acceleration to 5~GeV (b) and downstream of a single-stage bunch compressor (c) and current distribution (d) at the injector exit (``injector") and downstream of the bunch compressor (``after BC").}
\end{figure}

The noted reduction in longitudinal emittance combined with longitudinal bunch compression could further enhance the luminosity given the scaling ${\mathfrak{L}} \propto \sigma_z^{-1/2}$; see Eq.~\ref{eq:luminosity2}. In addition to improving luminosity, colliding short bunches also mitigate beamstrahlung-radiation losses thereby allowing the particles to experience extreme electromagnetic fields to probe non-perturbative quantum-electrodynamics effects~\cite{yakimenko-2019-a}. The photoinjector described in Sec.~\ref{sec:injetor} produces a final LPS with bunch length $\sigma_{z,e}=407$~\textmu{m}; see Fig.~\ref{fig:LPScompression}(a). Further accelerating the beam to 5~GeV [see Fig.~\ref{fig:LPScompression}(b)] and considering a single-stage bunch compressor (as implemented in the nominal ILC design downstream of the DR~\cite{latina-2010-a}) can reduce the bunch length to $\sigma'_{z}\simeq 23$~\textmu{m}; see Fig.~\ref{fig:LPScompression}(c,d). 
The simulations presented in Fig.~\ref{fig:LPScompression} were performed with a 1D single-particle model of the longitudinal beam dynamics. In the model, the linac accelerates the beam from 151~MeV to 5~GeV. The linac runs $15^{\circ}$ off-crest to impart the required correlated energy spread for maximum compression in a downstream bunch compressor. The bunch compressor is modeled by its longitudinal dispersion $R_{56}=14.9$~cm (in our convention $R_{56}>0$ corresponds to a chicane-like compressor). 

\section{conclusion}
In summary, we demonstrated a beamline composed of two cascaded cross-plane manipulations that could produce an electron beam with a final transverse-emittance partition comparable to the one attained downstream of the damping ring in the proposed ILC design. Additionally, our method produces  electron bunches with brightness  $\sim 2$ orders of magnitude higher than the ILC design. The enhanced brightness could further increase the luminosity by producing shorter bunches at the interaction point. Finally, the proposed scheme presents a substantial cost and complexity reduction compared to a damping ring. Although our focused was on demonstrating the application of the scheme to ILC-like parameters, the concept also be optimized for other LC designs. 

Yet, the integration of the proposed technique in future LC designs is contingent on the successful generation of spin-polarized beams from RF guns. Likewise, the method could also apply to positron beams pending the availability of low-emittance positron sources such as, e.g., recently proposed based on an electrostatic trap ~\cite{hessami-2022-a}, or relying on bremsstrahlung by impinging electron beams on thin targets~\cite{abbott-2016-a}. \\

Ultimately, the emittance-manipulation method discussed in this paper will require a vigorous R\&D program on sources of bright spin-polarized electron and positron beams to be deployed in a future LC design. Two complementary experiments aimed at testing the proposed concepts are currently in preparation at the Argonne Wakefield Accelerator (AWA)~\cite{xu-2022-a} and the Superconducting Test Facility (STF) at the High Energy Accelerator Research Organization (KEK)~\cite{kuriki-2018-a}. 
%
%
\section{acknowledgements}
We thank Dr. Eliana Gianfelice-Wendt (FNAL) for her suggestions regarding spin-dynamics simulations along with Drs. Chris Mayes (SLAC) and David Sagan (Cornell University) for their help with the spin-tracking feature of {\sc bmad}. We appreciate discussions with Dr. Zachary Liptak (Hiroshima University). PP is grateful to Dr. Spencer Gessner (SLAC) for sharing a draft  manuscript~\cite{hessami-2022-a} on bright positron sources. This research was initiated in the framework of the ``US-DOE-Japan cooperation in High-Energy Physics" program. This work was supported by the U.S. Department of Energy (DOE), Office of Science, under award No. DE-SC0018656 with Northern Illinois University (NIU) and contract No. DE-AC02-06CH11357 with Argonne National Laboratory (ANL).

The computing resources used for this research were provided on {\sc bebop},  a high-performance computing cluster operated by the Laboratory Computing Resource Center (LCRC) at ANL.

\bibliography{sample}

\begin{thebibliography}{48}%
\makeatletter
\providecommand \@ifxundefined [1]{%
 \@ifx{#1\undefined}
}%
\providecommand \@ifnum [1]{%
 \ifnum #1\expandafter \@firstoftwo
 \else \expandafter \@secondoftwo
 \fi
}%
\providecommand \@ifx [1]{%
 \ifx #1\expandafter \@firstoftwo
 \else \expandafter \@secondoftwo
 \fi
}%
\providecommand \natexlab [1]{#1}%
\providecommand \enquote  [1]{``#1''}%
\providecommand \bibnamefont  [1]{#1}%
\providecommand \bibfnamefont [1]{#1}%
\providecommand \citenamefont [1]{#1}%
\providecommand \href@noop [0]{\@secondoftwo}%
\providecommand \href [0]{\begingroup \@sanitize@url \@href}%
\providecommand \@href[1]{\@@startlink{#1}\@@href}%
\providecommand \@@href[1]{\endgroup#1\@@endlink}%
\providecommand \@sanitize@url [0]{\catcode `\\12\catcode `\$12\catcode
  `\&12\catcode `\#12\catcode `\^12\catcode `\_12\catcode `\%12\relax}%
\providecommand \@@startlink[1]{}%
\providecommand \@@endlink[0]{}%
\providecommand \url  [0]{\begingroup\@sanitize@url \@url }%
\providecommand \@url [1]{\endgroup\@href {#1}{\urlprefix }}%
\providecommand \urlprefix  [0]{URL }%
\providecommand \Eprint [0]{\href }%
\providecommand \doibase [0]{https://doi.org/}%
\providecommand \selectlanguage [0]{\@gobble}%
\providecommand \bibinfo  [0]{\@secondoftwo}%
\providecommand \bibfield  [0]{\@secondoftwo}%
\providecommand \translation [1]{[#1]}%
\providecommand \BibitemOpen [0]{}%
\providecommand \bibitemStop [0]{}%
\providecommand \bibitemNoStop [0]{.\EOS\space}%
\providecommand \EOS [0]{\spacefactor3000\relax}%
\providecommand \BibitemShut  [1]{\csname bibitem#1\endcsname}%
\let\auto@bib@innerbib\@empty
\bibitem [{\citenamefont {Ellis}\ and\ \citenamefont
  {Wilson}(2001)}]{ellis-2001}%
  \BibitemOpen
  \bibfield  {author} {\bibinfo {author} {\bibfnamefont {J.}~\bibnamefont
  {Ellis}}\ and\ \bibinfo {author} {\bibfnamefont {I.}~\bibnamefont {Wilson}},\
  }\bibfield  {title} {\bibinfo {title} {New physics with the compact linear
  collider},\ }\href {https://doi.org/10.1038/35053224} {\bibfield  {journal}
  {\bibinfo  {journal} {Nature}\ }\textbf {\bibinfo {volume} {409}},\ \bibinfo
  {pages} {431} (\bibinfo {year} {2001})}\BibitemShut {NoStop}%
\bibitem [{\citenamefont {Dugan}(2004)}]{dugan-2004-a}%
  \BibitemOpen
  \bibfield  {author} {\bibinfo {author} {\bibfnamefont {G.}~\bibnamefont
  {Dugan}},\ }\bibfield  {title} {\bibinfo {title} {Advanced accelerator system
  requirements for future linear colliders},\ }\href
  {https://doi.org/10.1063/1.1842533} {\bibfield  {journal} {\bibinfo
  {journal} {AIP Conference Proceedings}\ }\textbf {\bibinfo {volume} {737}},\
  \bibinfo {pages} {29} (\bibinfo {year} {2004})}\BibitemShut {NoStop}%
\bibitem [{\citenamefont {Yokoya}(2001)}]{yokoya-2001-a}%
  \BibitemOpen
  \bibfield  {author} {\bibinfo {author} {\bibfnamefont {K.}~\bibnamefont
  {Yokoya}},\ }\bibfield  {title} {\bibinfo {title} {Beam-beam interaction in
  linear collider},\ }\href {https://doi.org/10.1063/1.1420416} {\bibfield
  {journal} {\bibinfo  {journal} {AIP Conference Proceedings}\ }\textbf
  {\bibinfo {volume} {592}},\ \bibinfo {pages} {185} (\bibinfo {year}
  {2001})}\BibitemShut {NoStop}%
\bibitem [{\citenamefont {Krasilnikov}\ \emph {et~al.}(2012)\citenamefont
  {Krasilnikov}, \citenamefont {Stephan}, \citenamefont {Asova}, \citenamefont
  {Grabosch}, \citenamefont {Gro\ss{}}, \citenamefont {Hakobyan}, \citenamefont
  {Isaev}, \citenamefont {Ivanisenko}, \citenamefont {Jachmann}, \citenamefont
  {Khojoyan}, \citenamefont {Klemz}, \citenamefont {K\"ohler}, \citenamefont
  {Mahgoub}, \citenamefont {Malyutin}, \citenamefont {Nozdrin}, \citenamefont
  {Oppelt}, \citenamefont {Otevrel}, \citenamefont {Petrosyan}, \citenamefont
  {Rimjaem}, \citenamefont {Shapovalov}, \citenamefont {Vashchenko},
  \citenamefont {Weidinger}, \citenamefont {Wenndorff}, \citenamefont
  {Fl\"ottmann}, \citenamefont {Hoffmann}, \citenamefont {Lederer},
  \citenamefont {Schlarb}, \citenamefont {Schreiber}, \citenamefont {Templin},
  \citenamefont {Will}, \citenamefont {Paramonov},\ and\ \citenamefont
  {Richter}}]{krasilnikov-2012-a}%
  \BibitemOpen
  \bibfield  {author} {\bibinfo {author} {\bibfnamefont {M.}~\bibnamefont
  {Krasilnikov}}, \bibinfo {author} {\bibfnamefont {F.}~\bibnamefont
  {Stephan}}, \bibinfo {author} {\bibfnamefont {G.}~\bibnamefont {Asova}},
  \bibinfo {author} {\bibfnamefont {H.-J.}\ \bibnamefont {Grabosch}}, \bibinfo
  {author} {\bibfnamefont {M.}~\bibnamefont {Gro\ss{}}}, \bibinfo {author}
  {\bibfnamefont {L.}~\bibnamefont {Hakobyan}}, \bibinfo {author}
  {\bibfnamefont {I.}~\bibnamefont {Isaev}}, \bibinfo {author} {\bibfnamefont
  {Y.}~\bibnamefont {Ivanisenko}}, \bibinfo {author} {\bibfnamefont
  {L.}~\bibnamefont {Jachmann}}, \bibinfo {author} {\bibfnamefont
  {M.}~\bibnamefont {Khojoyan}}, \bibinfo {author} {\bibfnamefont
  {G.}~\bibnamefont {Klemz}}, \bibinfo {author} {\bibfnamefont
  {W.}~\bibnamefont {K\"ohler}}, \bibinfo {author} {\bibfnamefont
  {M.}~\bibnamefont {Mahgoub}}, \bibinfo {author} {\bibfnamefont
  {D.}~\bibnamefont {Malyutin}}, \bibinfo {author} {\bibfnamefont
  {M.}~\bibnamefont {Nozdrin}}, \bibinfo {author} {\bibfnamefont
  {A.}~\bibnamefont {Oppelt}}, \bibinfo {author} {\bibfnamefont
  {M.}~\bibnamefont {Otevrel}}, \bibinfo {author} {\bibfnamefont
  {B.}~\bibnamefont {Petrosyan}}, \bibinfo {author} {\bibfnamefont
  {S.}~\bibnamefont {Rimjaem}}, \bibinfo {author} {\bibfnamefont
  {A.}~\bibnamefont {Shapovalov}}, \bibinfo {author} {\bibfnamefont
  {G.}~\bibnamefont {Vashchenko}}, \bibinfo {author} {\bibfnamefont
  {S.}~\bibnamefont {Weidinger}}, \bibinfo {author} {\bibfnamefont
  {R.}~\bibnamefont {Wenndorff}}, \bibinfo {author} {\bibfnamefont
  {K.}~\bibnamefont {Fl\"ottmann}}, \bibinfo {author} {\bibfnamefont
  {M.}~\bibnamefont {Hoffmann}}, \bibinfo {author} {\bibfnamefont
  {S.}~\bibnamefont {Lederer}}, \bibinfo {author} {\bibfnamefont
  {H.}~\bibnamefont {Schlarb}}, \bibinfo {author} {\bibfnamefont
  {S.}~\bibnamefont {Schreiber}}, \bibinfo {author} {\bibfnamefont
  {I.}~\bibnamefont {Templin}}, \bibinfo {author} {\bibfnamefont
  {I.}~\bibnamefont {Will}}, \bibinfo {author} {\bibfnamefont {V.}~\bibnamefont
  {Paramonov}},\ and\ \bibinfo {author} {\bibfnamefont {D.}~\bibnamefont
  {Richter}},\ }\bibfield  {title} {\bibinfo {title} {Experimentally minimized
  beam emittance from an $l$-band photoinjector},\ }\href
  {https://doi.org/10.1103/PhysRevSTAB.15.100701} {\bibfield  {journal}
  {\bibinfo  {journal} {Phys. Rev. ST Accel. Beams}\ }\textbf {\bibinfo
  {volume} {15}},\ \bibinfo {pages} {100701} (\bibinfo {year}
  {2012})}\BibitemShut {NoStop}%
\bibitem [{\citenamefont {Brinkmann}\ \emph {et~al.}(2001)\citenamefont
  {Brinkmann}, \citenamefont {Derbenev},\ and\ \citenamefont
  {Fl\"ottmann}}]{brinkmann-2001-a}%
  \BibitemOpen
  \bibfield  {author} {\bibinfo {author} {\bibfnamefont {R.}~\bibnamefont
  {Brinkmann}}, \bibinfo {author} {\bibfnamefont {Y.}~\bibnamefont
  {Derbenev}},\ and\ \bibinfo {author} {\bibfnamefont {K.}~\bibnamefont
  {Fl\"ottmann}},\ }\bibfield  {title} {\bibinfo {title} {A low emittance,
  flat-beam electron source for linear colliders},\ }\href
  {https://doi.org/10.1103/PhysRevSTAB.4.053501} {\bibfield  {journal}
  {\bibinfo  {journal} {Phys. Rev. ST Accel. Beams}\ }\textbf {\bibinfo
  {volume} {4}},\ \bibinfo {pages} {053501} (\bibinfo {year}
  {2001})}\BibitemShut {NoStop}%
\bibitem [{\citenamefont {Phinney}\ \emph {et~al.}(2007)\citenamefont
  {Phinney}, \citenamefont {Toge},\ and\ \citenamefont
  {Walker}}]{phinney-2007-a}%
  \BibitemOpen
  \bibfield  {author} {\bibinfo {author} {\bibfnamefont {N.}~\bibnamefont
  {Phinney}}, \bibinfo {author} {\bibfnamefont {N.}~\bibnamefont {Toge}},\ and\
  \bibinfo {author} {\bibfnamefont {N.}~\bibnamefont {Walker}},\ }\href
  {https://doi.org/10.48550/ARXIV.0712.2361} {\bibinfo {title} {{ILC} reference
  design report volume 3 - accelerator}} (\bibinfo {year} {2007})\BibitemShut
  {NoStop}%
\bibitem [{\citenamefont {CLIC}\ \emph {et~al.}(2018)\citenamefont {CLIC},
  \citenamefont {Charles}, \citenamefont {Giansiracusa}, \citenamefont {Lucas},
  \citenamefont {Rassool}, \citenamefont {Volpi}, \citenamefont {Balazs},
  \citenamefont {Afanaciev}, \citenamefont {Makarenko}, \citenamefont
  {Patapenka} \emph {et~al.}}]{clic-2018}%
  \BibitemOpen
  \bibfield  {author} {\bibinfo {author} {\bibfnamefont {T.}~\bibnamefont
  {CLIC}}, \bibinfo {author} {\bibfnamefont {T.}~\bibnamefont {Charles}},
  \bibinfo {author} {\bibfnamefont {P.}~\bibnamefont {Giansiracusa}}, \bibinfo
  {author} {\bibfnamefont {T.}~\bibnamefont {Lucas}}, \bibinfo {author}
  {\bibfnamefont {R.}~\bibnamefont {Rassool}}, \bibinfo {author} {\bibfnamefont
  {M.}~\bibnamefont {Volpi}}, \bibinfo {author} {\bibfnamefont
  {C.}~\bibnamefont {Balazs}}, \bibinfo {author} {\bibfnamefont
  {K.}~\bibnamefont {Afanaciev}}, \bibinfo {author} {\bibfnamefont
  {V.}~\bibnamefont {Makarenko}}, \bibinfo {author} {\bibfnamefont
  {A.}~\bibnamefont {Patapenka}}, \emph {et~al.},\ }\bibfield  {title}
  {\bibinfo {title} {The compact linear collider (clic)-2018 summary report},\
  }\href {https://arxiv.org/abs/1812.06018} {\bibfield  {journal} {\bibinfo
  {journal} {arXiv preprint arXiv:1812.06018}\ } (\bibinfo {year}
  {2018})}\BibitemShut {NoStop}%
\bibitem [{\citenamefont {Cornacchia}\ and\ \citenamefont
  {Emma}(2002)}]{cornacchia-2002-a}%
  \BibitemOpen
  \bibfield  {author} {\bibinfo {author} {\bibfnamefont {M.}~\bibnamefont
  {Cornacchia}}\ and\ \bibinfo {author} {\bibfnamefont {P.}~\bibnamefont
  {Emma}},\ }\bibfield  {title} {\bibinfo {title} {Transverse to longitudinal
  emittance exchange},\ }\href {https://doi.org/10.1103/PhysRevSTAB.5.084001}
  {\bibfield  {journal} {\bibinfo  {journal} {Phys. Rev. ST Accel. Beams}\
  }\textbf {\bibinfo {volume} {5}},\ \bibinfo {pages} {084001} (\bibinfo {year}
  {2002})}\BibitemShut {NoStop}%
\bibitem [{\citenamefont {Emma}\ \emph {et~al.}(2006)\citenamefont {Emma},
  \citenamefont {Huang}, \citenamefont {Kim},\ and\ \citenamefont
  {Piot}}]{emma-2006-a}%
  \BibitemOpen
  \bibfield  {author} {\bibinfo {author} {\bibfnamefont {P.}~\bibnamefont
  {Emma}}, \bibinfo {author} {\bibfnamefont {Z.}~\bibnamefont {Huang}},
  \bibinfo {author} {\bibfnamefont {K.-J.}\ \bibnamefont {Kim}},\ and\ \bibinfo
  {author} {\bibfnamefont {P.}~\bibnamefont {Piot}},\ }\bibfield  {title}
  {\bibinfo {title} {Transverse-to-longitudinal emittance exchange to improve
  performance of high-gain free-electron lasers},\ }\href
  {https://doi.org/10.1103/PhysRevSTAB.9.100702} {\bibfield  {journal}
  {\bibinfo  {journal} {Phys. Rev. ST Accel. Beams}\ }\textbf {\bibinfo
  {volume} {9}},\ \bibinfo {pages} {100702} (\bibinfo {year}
  {2006})}\BibitemShut {NoStop}%
\bibitem [{\citenamefont {Piot}\ \emph {et~al.}(2006)\citenamefont {Piot},
  \citenamefont {Sun},\ and\ \citenamefont {Kim}}]{piot-2006-a}%
  \BibitemOpen
  \bibfield  {author} {\bibinfo {author} {\bibfnamefont {P.}~\bibnamefont
  {Piot}}, \bibinfo {author} {\bibfnamefont {Y.-E.}\ \bibnamefont {Sun}},\ and\
  \bibinfo {author} {\bibfnamefont {K.-J.}\ \bibnamefont {Kim}},\ }\bibfield
  {title} {\bibinfo {title} {Photoinjector generation of a flat electron beam
  with transverse emittance ratio of 100},\ }\href
  {https://doi.org/10.1103/PhysRevSTAB.9.031001} {\bibfield  {journal}
  {\bibinfo  {journal} {Phys. Rev. ST Accel. Beams}\ }\textbf {\bibinfo
  {volume} {9}},\ \bibinfo {pages} {031001} (\bibinfo {year}
  {2006})}\BibitemShut {NoStop}%
\bibitem [{\citenamefont {Ruan}\ \emph {et~al.}(2011)\citenamefont {Ruan},
  \citenamefont {Johnson}, \citenamefont {Lumpkin}, \citenamefont
  {Thurman-Keup}, \citenamefont {Edwards}, \citenamefont {Fliller},
  \citenamefont {Koeth},\ and\ \citenamefont {Sun}}]{ruan-2011-a}%
  \BibitemOpen
  \bibfield  {author} {\bibinfo {author} {\bibfnamefont {J.}~\bibnamefont
  {Ruan}}, \bibinfo {author} {\bibfnamefont {A.~S.}\ \bibnamefont {Johnson}},
  \bibinfo {author} {\bibfnamefont {A.~H.}\ \bibnamefont {Lumpkin}}, \bibinfo
  {author} {\bibfnamefont {R.}~\bibnamefont {Thurman-Keup}}, \bibinfo {author}
  {\bibfnamefont {H.}~\bibnamefont {Edwards}}, \bibinfo {author} {\bibfnamefont
  {R.~P.}\ \bibnamefont {Fliller}}, \bibinfo {author} {\bibfnamefont {T.~W.}\
  \bibnamefont {Koeth}},\ and\ \bibinfo {author} {\bibfnamefont {Y.-E.}\
  \bibnamefont {Sun}},\ }\bibfield  {title} {\bibinfo {title} {First
  observation of the exchange of transverse and longitudinal emittances},\
  }\href {https://doi.org/10.1103/PhysRevLett.106.244801} {\bibfield  {journal}
  {\bibinfo  {journal} {Phys. Rev. Lett.}\ }\textbf {\bibinfo {volume} {106}},\
  \bibinfo {pages} {244801} (\bibinfo {year} {2011})}\BibitemShut {NoStop}%
\bibitem [{\citenamefont {Sun}\ \emph {et~al.}(2010)\citenamefont {Sun},
  \citenamefont {Piot}, \citenamefont {Johnson}, \citenamefont {Lumpkin},
  \citenamefont {Maxwell}, \citenamefont {Ruan},\ and\ \citenamefont
  {Thurman-Keup}}]{sun-2010-a}%
  \BibitemOpen
  \bibfield  {author} {\bibinfo {author} {\bibfnamefont {Y.-E.}\ \bibnamefont
  {Sun}}, \bibinfo {author} {\bibfnamefont {P.}~\bibnamefont {Piot}}, \bibinfo
  {author} {\bibfnamefont {A.}~\bibnamefont {Johnson}}, \bibinfo {author}
  {\bibfnamefont {A.~H.}\ \bibnamefont {Lumpkin}}, \bibinfo {author}
  {\bibfnamefont {T.~J.}\ \bibnamefont {Maxwell}}, \bibinfo {author}
  {\bibfnamefont {J.}~\bibnamefont {Ruan}},\ and\ \bibinfo {author}
  {\bibfnamefont {R.}~\bibnamefont {Thurman-Keup}},\ }\bibfield  {title}
  {\bibinfo {title} {Tunable subpicosecond electron-bunch-train generation
  using a transverse-to-longitudinal phase-space exchange technique},\ }\href
  {https://doi.org/10.1103/PhysRevLett.105.234801} {\bibfield  {journal}
  {\bibinfo  {journal} {Phys. Rev. Lett.}\ }\textbf {\bibinfo {volume} {105}},\
  \bibinfo {pages} {234801} (\bibinfo {year} {2010})}\BibitemShut {NoStop}%
\bibitem [{\citenamefont {Groening}\ \emph {et~al.}(2014)\citenamefont
  {Groening}, \citenamefont {Maier}, \citenamefont {Xiao}, \citenamefont
  {Dahl}, \citenamefont {Gerhard}, \citenamefont {Kester}, \citenamefont
  {Mickat}, \citenamefont {Vormann}, \citenamefont {Vossberg},\ and\
  \citenamefont {Chung}}]{groening-2014-a}%
  \BibitemOpen
  \bibfield  {author} {\bibinfo {author} {\bibfnamefont {L.}~\bibnamefont
  {Groening}}, \bibinfo {author} {\bibfnamefont {M.}~\bibnamefont {Maier}},
  \bibinfo {author} {\bibfnamefont {C.}~\bibnamefont {Xiao}}, \bibinfo {author}
  {\bibfnamefont {L.}~\bibnamefont {Dahl}}, \bibinfo {author} {\bibfnamefont
  {P.}~\bibnamefont {Gerhard}}, \bibinfo {author} {\bibfnamefont {O.~K.}\
  \bibnamefont {Kester}}, \bibinfo {author} {\bibfnamefont {S.}~\bibnamefont
  {Mickat}}, \bibinfo {author} {\bibfnamefont {H.}~\bibnamefont {Vormann}},
  \bibinfo {author} {\bibfnamefont {M.}~\bibnamefont {Vossberg}},\ and\
  \bibinfo {author} {\bibfnamefont {M.}~\bibnamefont {Chung}},\ }\bibfield
  {title} {\bibinfo {title} {Experimental proof of adjustable single-knob ion
  beam emittance partitioning},\ }\href
  {https://doi.org/10.1103/PhysRevLett.113.264802} {\bibfield  {journal}
  {\bibinfo  {journal} {Phys. Rev. Lett.}\ }\textbf {\bibinfo {volume} {113}},\
  \bibinfo {pages} {264802} (\bibinfo {year} {2014})}\BibitemShut {NoStop}%
\bibitem [{\citenamefont {Ha}\ \emph {et~al.}(2016)\citenamefont {Ha},
  \citenamefont {Cho}, \citenamefont {Gai}, \citenamefont {Kim}, \citenamefont
  {Namkung},\ and\ \citenamefont {Power}}]{ha-2016}%
  \BibitemOpen
  \bibfield  {author} {\bibinfo {author} {\bibfnamefont {G.}~\bibnamefont
  {Ha}}, \bibinfo {author} {\bibfnamefont {M.}~\bibnamefont {Cho}}, \bibinfo
  {author} {\bibfnamefont {W.}~\bibnamefont {Gai}}, \bibinfo {author}
  {\bibfnamefont {K.-J.}\ \bibnamefont {Kim}}, \bibinfo {author} {\bibfnamefont
  {W.}~\bibnamefont {Namkung}},\ and\ \bibinfo {author} {\bibfnamefont
  {J.}~\bibnamefont {Power}},\ }\bibfield  {title} {\bibinfo {title}
  {Perturbation-minimized triangular bunch for high-transformer ratio using a
  double dogleg emittance exchange beam line},\ }\href
  {https://journals.aps.org/prab/abstract/10.1103/PhysRevAccelBeams.19.121301}
  {\bibfield  {journal} {\bibinfo  {journal} {Phys. Rev. Accel. Beams}\
  }\textbf {\bibinfo {volume} {19}},\ \bibinfo {pages} {121301} (\bibinfo
  {year} {2016})}\BibitemShut {NoStop}%
\bibitem [{\citenamefont {Yampolsky}\ \emph {et~al.}(2010)\citenamefont
  {Yampolsky}, \citenamefont {Carlsten}, \citenamefont {Ryne}, \citenamefont
  {Bishofberger}, \citenamefont {Russell},\ and\ \citenamefont
  {Dragt}}]{yampolsky-2010-a}%
  \BibitemOpen
  \bibfield  {author} {\bibinfo {author} {\bibfnamefont {N.}~\bibnamefont
  {Yampolsky}}, \bibinfo {author} {\bibfnamefont {B.}~\bibnamefont {Carlsten}},
  \bibinfo {author} {\bibfnamefont {R.}~\bibnamefont {Ryne}}, \bibinfo {author}
  {\bibfnamefont {K.}~\bibnamefont {Bishofberger}}, \bibinfo {author}
  {\bibfnamefont {S.}~\bibnamefont {Russell}},\ and\ \bibinfo {author}
  {\bibfnamefont {A.}~\bibnamefont {Dragt}},\ }\href
  {https://arxiv.org/abs/1010.1558} {\bibinfo {title} {Controlling
  electron-beam emittance partitioning for future x-ray light sources}}
  (\bibinfo {year} {2010}),\ \Eprint {https://arxiv.org/abs/1010.1558}
  {arXiv:1010.1558 [physics.acc-ph]} \BibitemShut {NoStop}%
\bibitem [{\citenamefont {Carlsten}\ \emph {et~al.}(2011)\citenamefont
  {Carlsten}, \citenamefont {Bishofberger}, \citenamefont {Duffy},
  \citenamefont {Russell}, \citenamefont {Ryne}, \citenamefont {Yampolsky},\
  and\ \citenamefont {Dragt}}]{carlsten-2011-a}%
  \BibitemOpen
  \bibfield  {author} {\bibinfo {author} {\bibfnamefont {B.~E.}\ \bibnamefont
  {Carlsten}}, \bibinfo {author} {\bibfnamefont {K.~A.}\ \bibnamefont
  {Bishofberger}}, \bibinfo {author} {\bibfnamefont {L.~D.}\ \bibnamefont
  {Duffy}}, \bibinfo {author} {\bibfnamefont {S.~J.}\ \bibnamefont {Russell}},
  \bibinfo {author} {\bibfnamefont {R.~D.}\ \bibnamefont {Ryne}}, \bibinfo
  {author} {\bibfnamefont {N.~A.}\ \bibnamefont {Yampolsky}},\ and\ \bibinfo
  {author} {\bibfnamefont {A.~J.}\ \bibnamefont {Dragt}},\ }\bibfield  {title}
  {\bibinfo {title} {Arbitrary emittance partitioning between any two
  dimensions for electron beams},\ }\href
  {https://doi.org/10.1103/PhysRevSTAB.14.050706} {\bibfield  {journal}
  {\bibinfo  {journal} {Phys. Rev. ST Accel. Beams}\ }\textbf {\bibinfo
  {volume} {14}},\ \bibinfo {pages} {050706} (\bibinfo {year}
  {2011})}\BibitemShut {NoStop}%
\bibitem [{\citenamefont {Duffy}\ and\ \citenamefont
  {Dragt}(2016)}]{duffy-2016-a}%
  \BibitemOpen
  \bibfield  {author} {\bibinfo {author} {\bibfnamefont {L.~D.}\ \bibnamefont
  {Duffy}}\ and\ \bibinfo {author} {\bibfnamefont {A.~J.}\ \bibnamefont
  {Dragt}},\ }\bibfield  {title} {\bibinfo {title} {Chapter one - utilizing the
  eigen-emittance concept for bright electron beams},\ }in\ \href
  {https://doi.org/https://doi.org/10.1016/bs.aiep.2015.11.001} {\emph
  {\bibinfo {booktitle} {Advances in Imaging and Electron Physics}}},\ Vol.\
  \bibinfo {volume} {193},\ \bibinfo {editor} {edited by\ \bibinfo {editor}
  {\bibfnamefont {P.~W.}\ \bibnamefont {Hawkes}}}\ (\bibinfo  {publisher}
  {Elsevier},\ \bibinfo {year} {2016})\ pp.\ \bibinfo {pages}
  {1--44}\BibitemShut {NoStop}%
\bibitem [{\citenamefont {Kim}(2003)}]{kim-2003-a}%
  \BibitemOpen
  \bibfield  {author} {\bibinfo {author} {\bibfnamefont {K.-J.}\ \bibnamefont
  {Kim}},\ }\bibfield  {title} {\bibinfo {title} {Round-to-flat transformation
  of angular-momentum-dominated beams},\ }\href
  {https://doi.org/10.1103/PhysRevSTAB.6.104002} {\bibfield  {journal}
  {\bibinfo  {journal} {Phys. Rev. ST Accel. Beams}\ }\textbf {\bibinfo
  {volume} {6}},\ \bibinfo {pages} {104002} (\bibinfo {year}
  {2003})}\BibitemShut {NoStop}%
\bibitem [{\citenamefont {Rosenzweig}\ and\ \citenamefont
  {Serafini}(1994)}]{rosen-1994-a}%
  \BibitemOpen
  \bibfield  {author} {\bibinfo {author} {\bibfnamefont {J.}~\bibnamefont
  {Rosenzweig}}\ and\ \bibinfo {author} {\bibfnamefont {L.}~\bibnamefont
  {Serafini}},\ }\bibfield  {title} {\bibinfo {title} {Transverse particle
  motion in radio-frequency linear accelerators},\ }\href
  {https://doi.org/10.1103/PhysRevE.49.1599} {\bibfield  {journal} {\bibinfo
  {journal} {Phys. Rev. E}\ }\textbf {\bibinfo {volume} {49}},\ \bibinfo
  {pages} {1599} (\bibinfo {year} {1994})}\BibitemShut {NoStop}%
\bibitem [{\citenamefont {Sun}\ \emph {et~al.}(2004)\citenamefont {Sun},
  \citenamefont {Piot}, \citenamefont {Kim}, \citenamefont {Barov},
  \citenamefont {Lidia}, \citenamefont {Santucci}, \citenamefont {Tikhoplav},\
  and\ \citenamefont {Wennerberg}}]{sun-2004-a}%
  \BibitemOpen
  \bibfield  {author} {\bibinfo {author} {\bibfnamefont {Y.}~\bibnamefont
  {Sun}}, \bibinfo {author} {\bibfnamefont {P.}~\bibnamefont {Piot}}, \bibinfo
  {author} {\bibfnamefont {K.-J.}\ \bibnamefont {Kim}}, \bibinfo {author}
  {\bibfnamefont {N.}~\bibnamefont {Barov}}, \bibinfo {author} {\bibfnamefont
  {S.}~\bibnamefont {Lidia}}, \bibinfo {author} {\bibfnamefont
  {J.}~\bibnamefont {Santucci}}, \bibinfo {author} {\bibfnamefont
  {R.}~\bibnamefont {Tikhoplav}},\ and\ \bibinfo {author} {\bibfnamefont
  {J.}~\bibnamefont {Wennerberg}},\ }\bibfield  {title} {\bibinfo {title}
  {Generation of angular-momentum-dominated electron beams from a
  photoinjector},\ }\href {https://doi.org/10.1103/PhysRevSTAB.7.123501}
  {\bibfield  {journal} {\bibinfo  {journal} {Phys. Rev. ST Accel. Beams}\
  }\textbf {\bibinfo {volume} {7}},\ \bibinfo {pages} {123501} (\bibinfo {year}
  {2004})}\BibitemShut {NoStop}%
\bibitem [{\citenamefont {Burov}\ \emph {et~al.}(2002)\citenamefont {Burov},
  \citenamefont {Nagaitsev},\ and\ \citenamefont {Derbenev}}]{burov-2002-a}%
  \BibitemOpen
  \bibfield  {author} {\bibinfo {author} {\bibfnamefont {A.}~\bibnamefont
  {Burov}}, \bibinfo {author} {\bibfnamefont {S.}~\bibnamefont {Nagaitsev}},\
  and\ \bibinfo {author} {\bibfnamefont {Y.}~\bibnamefont {Derbenev}},\
  }\bibfield  {title} {\bibinfo {title} {Circular modes, beam adapters, and
  their applications in beam optics},\ }\href
  {https://doi.org/10.1103/PhysRevE.66.016503} {\bibfield  {journal} {\bibinfo
  {journal} {Phys. Rev. E}\ }\textbf {\bibinfo {volume} {66}},\ \bibinfo
  {pages} {016503} (\bibinfo {year} {2002})}\BibitemShut {NoStop}%
\bibitem [{\citenamefont {Xu}\ \emph {et~al.}(2022)\citenamefont {Xu},
  \citenamefont {Doran}, \citenamefont {Liu}, \citenamefont {Piot},
  \citenamefont {Power}, \citenamefont {Whiteford},\ and\ \citenamefont
  {Wisniewski}}]{xu-2022-a}%
  \BibitemOpen
  \bibfield  {author} {\bibinfo {author} {\bibfnamefont {T.}~\bibnamefont
  {Xu}}, \bibinfo {author} {\bibfnamefont {D.~S.}\ \bibnamefont {Doran}},
  \bibinfo {author} {\bibfnamefont {W.}~\bibnamefont {Liu}}, \bibinfo {author}
  {\bibfnamefont {P.}~\bibnamefont {Piot}}, \bibinfo {author} {\bibfnamefont
  {J.~G.}\ \bibnamefont {Power}}, \bibinfo {author} {\bibfnamefont
  {C.}~\bibnamefont {Whiteford}},\ and\ \bibinfo {author} {\bibfnamefont
  {E.}~\bibnamefont {Wisniewski}},\ }\bibfield  {title} {\bibinfo {title}
  {Demonstration of eigen-to-projected emittance mapping for an ellipsoidal
  electron bunch},\ }\href
  {https://doi.org/10.1103/PhysRevAccelBeams.25.044001} {\bibfield  {journal}
  {\bibinfo  {journal} {Phys. Rev. Accel. Beams}\ }\textbf {\bibinfo {volume}
  {25}},\ \bibinfo {pages} {044001} (\bibinfo {year} {2022})}\BibitemShut
  {NoStop}%
\bibitem [{\citenamefont {Nanni}\ and\ \citenamefont
  {Graves}(2015)}]{nanni2015}%
  \BibitemOpen
  \bibfield  {author} {\bibinfo {author} {\bibfnamefont {E.~A.}\ \bibnamefont
  {Nanni}}\ and\ \bibinfo {author} {\bibfnamefont {W.~S.}\ \bibnamefont
  {Graves}},\ }\bibfield  {title} {\bibinfo {title} {Aberration corrected
  emittance exchange},\ }\href {https://doi.org/10.1103/PhysRevSTAB.18.084401}
  {\bibfield  {journal} {\bibinfo  {journal} {Phys. Rev. ST Accel. Beams}\
  }\textbf {\bibinfo {volume} {18}},\ \bibinfo {pages} {084401} (\bibinfo
  {year} {2015})}\BibitemShut {NoStop}%
\bibitem [{\citenamefont {Ha}\ \emph {et~al.}(2017)\citenamefont {Ha},
  \citenamefont {Cho}, \citenamefont {Namkung}, \citenamefont {Power},
  \citenamefont {Doran}, \citenamefont {Wisniewski}, \citenamefont {Conde},
  \citenamefont {Gai}, \citenamefont {Liu}, \citenamefont {Whiteford},
  \citenamefont {Gao}, \citenamefont {Kim}, \citenamefont {Zholents},
  \citenamefont {Sun}, \citenamefont {Jing},\ and\ \citenamefont
  {Piot}}]{ha-2017-a}%
  \BibitemOpen
  \bibfield  {author} {\bibinfo {author} {\bibfnamefont {G.}~\bibnamefont
  {Ha}}, \bibinfo {author} {\bibfnamefont {M.~H.}\ \bibnamefont {Cho}},
  \bibinfo {author} {\bibfnamefont {W.}~\bibnamefont {Namkung}}, \bibinfo
  {author} {\bibfnamefont {J.~G.}\ \bibnamefont {Power}}, \bibinfo {author}
  {\bibfnamefont {D.~S.}\ \bibnamefont {Doran}}, \bibinfo {author}
  {\bibfnamefont {E.~E.}\ \bibnamefont {Wisniewski}}, \bibinfo {author}
  {\bibfnamefont {M.}~\bibnamefont {Conde}}, \bibinfo {author} {\bibfnamefont
  {W.}~\bibnamefont {Gai}}, \bibinfo {author} {\bibfnamefont {W.}~\bibnamefont
  {Liu}}, \bibinfo {author} {\bibfnamefont {C.}~\bibnamefont {Whiteford}},
  \bibinfo {author} {\bibfnamefont {Q.}~\bibnamefont {Gao}}, \bibinfo {author}
  {\bibfnamefont {K.-J.}\ \bibnamefont {Kim}}, \bibinfo {author} {\bibfnamefont
  {A.}~\bibnamefont {Zholents}}, \bibinfo {author} {\bibfnamefont {Y.-E.}\
  \bibnamefont {Sun}}, \bibinfo {author} {\bibfnamefont {C.}~\bibnamefont
  {Jing}},\ and\ \bibinfo {author} {\bibfnamefont {P.}~\bibnamefont {Piot}},\
  }\bibfield  {title} {\bibinfo {title} {Precision control of the electron
  longitudinal bunch shape using an emittance-exchange beam line},\ }\href
  {https://doi.org/10.1103/PhysRevLett.118.104801} {\bibfield  {journal}
  {\bibinfo  {journal} {Phys. Rev. Lett.}\ }\textbf {\bibinfo {volume} {118}},\
  \bibinfo {pages} {104801} (\bibinfo {year} {2017})}\BibitemShut {NoStop}%
\bibitem [{\citenamefont {Zholents}\ and\ \citenamefont
  {Zolotorev}(2011)}]{zholents-2011-a}%
  \BibitemOpen
  \bibfield  {author} {\bibinfo {author} {\bibfnamefont {A.~A.}\ \bibnamefont
  {Zholents}}\ and\ \bibinfo {author} {\bibfnamefont {M.~S.}\ \bibnamefont
  {Zolotorev}},\ }\href {https://doi.org/https://doi.org/10.2172/1015557}
  {\emph {\bibinfo {title} {A new type of bunch compressor and seeding of short
  wave length coherent radiation}}},\ \bibinfo {type} {Tech. Rep.}\ \bibinfo
  {number} {ANL/APS/LS-327}\ (\bibinfo  {institution} {Argonne National
  Laboratory},\ \bibinfo {year} {2011})\BibitemShut {NoStop}%
\bibitem [{\citenamefont {Xiang}\ and\ \citenamefont
  {Chao}(2011)}]{xiang-2011-a}%
  \BibitemOpen
  \bibfield  {author} {\bibinfo {author} {\bibfnamefont {D.}~\bibnamefont
  {Xiang}}\ and\ \bibinfo {author} {\bibfnamefont {A.}~\bibnamefont {Chao}},\
  }\bibfield  {title} {\bibinfo {title} {Emittance and phase space exchange for
  advanced beam manipulation and diagnostics},\ }\href
  {https://doi.org/10.1103/PhysRevSTAB.14.114001} {\bibfield  {journal}
  {\bibinfo  {journal} {Phys. Rev. ST Accel. Beams}\ }\textbf {\bibinfo
  {volume} {14}},\ \bibinfo {pages} {114001} (\bibinfo {year}
  {2011})}\BibitemShut {NoStop}%
\bibitem [{\citenamefont {{Brown}}(1968)}]{brown-1975-a}%
  \BibitemOpen
  \bibfield  {author} {\bibinfo {author} {\bibfnamefont {K.~L.}\ \bibnamefont
  {{Brown}}},\ }\bibfield  {title} {\bibinfo {title} {{A First- and
  Second-Order Matrix Theory for the Design of Beam Transport Systems and
  Charged Particle Spectrometers}},\ }in\ \href@noop {} {\emph {\bibinfo
  {booktitle} {Advances in Particle Physics, Volume I}}},\ \bibinfo {editor}
  {edited by\ \bibinfo {editor} {\bibfnamefont {R.~L.}\ \bibnamefont {{Cool}}}\
  and\ \bibinfo {editor} {\bibfnamefont {R.~E.}\ \bibnamefont {{Marshak}}}}\
  (\bibinfo {year} {1968})\ p.~\bibinfo {pages} {71},\ \bibinfo {note} {also
  available as SLAC technical report SLAC-R-075, SLAC-75}\BibitemShut {NoStop}%
\bibitem [{\citenamefont {Qiang}\ \emph {et~al.}(2006)\citenamefont {Qiang},
  \citenamefont {Lidia}, \citenamefont {Ryne},\ and\ \citenamefont
  {Limborg-Deprey}}]{qiang-2006-a}%
  \BibitemOpen
  \bibfield  {author} {\bibinfo {author} {\bibfnamefont {J.}~\bibnamefont
  {Qiang}}, \bibinfo {author} {\bibfnamefont {S.}~\bibnamefont {Lidia}},
  \bibinfo {author} {\bibfnamefont {R.~D.}\ \bibnamefont {Ryne}},\ and\
  \bibinfo {author} {\bibfnamefont {C.}~\bibnamefont {Limborg-Deprey}},\
  }\bibfield  {title} {\bibinfo {title} {Three-dimensional quasistatic model
  for high brightness beam dynamics simulation},\ }\href
  {https://doi.org/10.1103/PhysRevSTAB.9.044204} {\bibfield  {journal}
  {\bibinfo  {journal} {Phys. Rev. ST Accel. Beams}\ }\textbf {\bibinfo
  {volume} {9}},\ \bibinfo {pages} {044204} (\bibinfo {year}
  {2006})}\BibitemShut {NoStop}%
\bibitem [{\citenamefont {Broemmelsiek}\ \emph {et~al.}(2018)\citenamefont
  {Broemmelsiek}, \citenamefont {Chase}, \citenamefont {Edstrom}, \citenamefont
  {Harms}, \citenamefont {Leibfritz}, \citenamefont {Nagaitsev}, \citenamefont
  {Pischalnikov}, \citenamefont {Romanov}, \citenamefont {Ruan}, \citenamefont
  {Schappert}, \citenamefont {Shiltsev}, \citenamefont {Thurman-Keup},\ and\
  \citenamefont {Valishev}}]{broemmelsiek-2018-a}%
  \BibitemOpen
  \bibfield  {author} {\bibinfo {author} {\bibfnamefont {D.}~\bibnamefont
  {Broemmelsiek}}, \bibinfo {author} {\bibfnamefont {B.}~\bibnamefont {Chase}},
  \bibinfo {author} {\bibfnamefont {D.}~\bibnamefont {Edstrom}}, \bibinfo
  {author} {\bibfnamefont {E.}~\bibnamefont {Harms}}, \bibinfo {author}
  {\bibfnamefont {J.}~\bibnamefont {Leibfritz}}, \bibinfo {author}
  {\bibfnamefont {S.}~\bibnamefont {Nagaitsev}}, \bibinfo {author}
  {\bibfnamefont {Y.}~\bibnamefont {Pischalnikov}}, \bibinfo {author}
  {\bibfnamefont {A.}~\bibnamefont {Romanov}}, \bibinfo {author} {\bibfnamefont
  {J.}~\bibnamefont {Ruan}}, \bibinfo {author} {\bibfnamefont {W.}~\bibnamefont
  {Schappert}}, \bibinfo {author} {\bibfnamefont {V.}~\bibnamefont {Shiltsev}},
  \bibinfo {author} {\bibfnamefont {R.}~\bibnamefont {Thurman-Keup}},\ and\
  \bibinfo {author} {\bibfnamefont {A.}~\bibnamefont {Valishev}},\ }\bibfield
  {title} {\bibinfo {title} {Record high-gradient {SRF} beam acceleration at
  fermilab},\ }\href {https://doi.org/10.1088/1367-2630/aaec57} {\bibfield
  {journal} {\bibinfo  {journal} {New Journal of Physics}\ }\textbf {\bibinfo
  {volume} {20}},\ \bibinfo {pages} {113018} (\bibinfo {year}
  {2018})}\BibitemShut {NoStop}%
\bibitem [{\citenamefont {Li}\ and\ \citenamefont
  {Lewellen}(2008)}]{li-2008-a}%
  \BibitemOpen
  \bibfield  {author} {\bibinfo {author} {\bibfnamefont {Y.}~\bibnamefont
  {Li}}\ and\ \bibinfo {author} {\bibfnamefont {J.~W.}\ \bibnamefont
  {Lewellen}},\ }\bibfield  {title} {\bibinfo {title} {Generating a
  quasiellipsoidal electron beam by {3D} laser-pulse shaping},\ }\href
  {https://doi.org/10.1103/PhysRevLett.100.074801} {\bibfield  {journal}
  {\bibinfo  {journal} {Phys. Rev. Lett.}\ }\textbf {\bibinfo {volume} {100}},\
  \bibinfo {pages} {074801} (\bibinfo {year} {2008})}\BibitemShut {NoStop}%
\bibitem [{\citenamefont {Li}\ \emph {et~al.}(2009)\citenamefont {Li},
  \citenamefont {Chemerisov},\ and\ \citenamefont {Lewellen}}]{li-2009-a}%
  \BibitemOpen
  \bibfield  {author} {\bibinfo {author} {\bibfnamefont {Y.}~\bibnamefont
  {Li}}, \bibinfo {author} {\bibfnamefont {S.}~\bibnamefont {Chemerisov}},\
  and\ \bibinfo {author} {\bibfnamefont {J.}~\bibnamefont {Lewellen}},\
  }\bibfield  {title} {\bibinfo {title} {Laser pulse shaping for generating
  uniform three-dimensional ellipsoidal electron beams},\ }\href
  {https://doi.org/10.1103/PhysRevSTAB.12.020702} {\bibfield  {journal}
  {\bibinfo  {journal} {Phys. Rev. ST Accel. Beams}\ }\textbf {\bibinfo
  {volume} {12}},\ \bibinfo {pages} {020702} (\bibinfo {year}
  {2009})}\BibitemShut {NoStop}%
\bibitem [{\citenamefont {Lapostolle}(1965)}]{lapostolle-1965-a}%
  \BibitemOpen
  \bibfield  {author} {\bibinfo {author} {\bibfnamefont {P.~M.}\ \bibnamefont
  {Lapostolle}},\ }\href {http://cds.cern.ch/record/355640} {\emph {\bibinfo
  {title} {{Effets de la charge d'espace dans un acc\'el\'erateur lin\'eaire
  \`a protons}}}},\ \bibinfo {type} {Tech. Rep.}\ \bibinfo {number}
  {CERN-AR-Int-SG-65-15}\ (\bibinfo  {institution} {CERN},\ \bibinfo {address}
  {Geneva},\ \bibinfo {year} {1965})\BibitemShut {NoStop}%
\bibitem [{\citenamefont {Luiten}\ \emph {et~al.}(2004)\citenamefont {Luiten},
  \citenamefont {van~der Geer}, \citenamefont {de~Loos}, \citenamefont
  {Kiewiet},\ and\ \citenamefont {van~der Wiel}}]{luiten-2004-a}%
  \BibitemOpen
  \bibfield  {author} {\bibinfo {author} {\bibfnamefont {O.~J.}\ \bibnamefont
  {Luiten}}, \bibinfo {author} {\bibfnamefont {S.~B.}\ \bibnamefont {van~der
  Geer}}, \bibinfo {author} {\bibfnamefont {M.~J.}\ \bibnamefont {de~Loos}},
  \bibinfo {author} {\bibfnamefont {F.~B.}\ \bibnamefont {Kiewiet}},\ and\
  \bibinfo {author} {\bibfnamefont {M.~J.}\ \bibnamefont {van~der Wiel}},\
  }\bibfield  {title} {\bibinfo {title} {How to realize uniform
  three-dimensional ellipsoidal electron bunches},\ }\href
  {https://doi.org/10.1103/PhysRevLett.93.094802} {\bibfield  {journal}
  {\bibinfo  {journal} {Phys. Rev. Lett.}\ }\textbf {\bibinfo {volume} {93}},\
  \bibinfo {pages} {094802} (\bibinfo {year} {2004})}\BibitemShut {NoStop}%
\bibitem [{\citenamefont {Bertucci}\ \emph {et~al.}(2019)\citenamefont
  {Bertucci}, \citenamefont {Bignami}, \citenamefont {Bosotti}, \citenamefont
  {Michelato}, \citenamefont {Monaco}, \citenamefont {Pagani}, \citenamefont
  {Paparella}, \citenamefont {Sertore}, \citenamefont {Maiano}, \citenamefont
  {Pierini},\ and\ \citenamefont {Chen}}]{bertucci-2019-a}%
  \BibitemOpen
  \bibfield  {author} {\bibinfo {author} {\bibfnamefont {M.}~\bibnamefont
  {Bertucci}}, \bibinfo {author} {\bibfnamefont {A.}~\bibnamefont {Bignami}},
  \bibinfo {author} {\bibfnamefont {A.}~\bibnamefont {Bosotti}}, \bibinfo
  {author} {\bibfnamefont {P.}~\bibnamefont {Michelato}}, \bibinfo {author}
  {\bibfnamefont {L.}~\bibnamefont {Monaco}}, \bibinfo {author} {\bibfnamefont
  {C.}~\bibnamefont {Pagani}}, \bibinfo {author} {\bibfnamefont
  {R.}~\bibnamefont {Paparella}}, \bibinfo {author} {\bibfnamefont
  {D.}~\bibnamefont {Sertore}}, \bibinfo {author} {\bibfnamefont
  {C.}~\bibnamefont {Maiano}}, \bibinfo {author} {\bibfnamefont
  {P.}~\bibnamefont {Pierini}},\ and\ \bibinfo {author} {\bibfnamefont
  {J.}~\bibnamefont {Chen}},\ }\bibfield  {title} {\bibinfo {title}
  {Performance analysis of the {E}uropean {X}-ray {F}ree {E}lectron {L}aser 3.9
  {GHz} superconducting cavities},\ }\href
  {https://doi.org/10.1103/PhysRevAccelBeams.22.082002} {\bibfield  {journal}
  {\bibinfo  {journal} {Phys. Rev. Accel. Beams}\ }\textbf {\bibinfo {volume}
  {22}},\ \bibinfo {pages} {082002} (\bibinfo {year} {2019})}\BibitemShut
  {NoStop}%
\bibitem [{\citenamefont {Borland}(2000)}]{borland-2000-a}%
  \BibitemOpen
  \bibfield  {author} {\bibinfo {author} {\bibfnamefont {M.}~\bibnamefont
  {Borland}},\ }\bibfield  {title} {\bibinfo {title} {{{ELEGANT}: A Flexible
  {SDDS}-Compliant Code for Accelerator Simulation}},\ }in\ \href
  {https://doi.org/10.2172/761286} {\emph {\bibinfo {booktitle} {{6th
  International Computational Accelerator Physics Conference (ICAP 2000)}}}}\
  (\bibinfo {year} {2000})\BibitemShut {NoStop}%
\bibitem [{\citenamefont {Omet}\ \emph {et~al.}(2018)\citenamefont {Omet} \emph
  {et~al.}}]{omet-2018-a}%
  \BibitemOpen
  \bibfield  {author} {\bibinfo {author} {\bibfnamefont {M.}~\bibnamefont
  {Omet}} \emph {et~al.},\ }\bibfield  {title} {\bibinfo {title} {{LLRF}
  {O}peration and {P}erformance at the {E}uropean {XFEL}},\ }in\ \href
  {https://doi.org/doi:10.18429/JACoW-IPAC2018-WEPAF051} {\emph {\bibinfo
  {booktitle} {Proc. 9th International Particle Accelerator Conference
  (IPAC'18), Vancouver, BC, Canada, April 29-May 4, 2018}}},\ \bibinfo {series
  and number} {\bibinfo {series} {International Particle Accelerator
  Conference}\ No.~\bibinfo {number} {9}}\ (\bibinfo  {publisher} {JACoW
  Publishing},\ \bibinfo {address} {Geneva, Switzerland},\ \bibinfo {year}
  {2018})\ pp.\ \bibinfo {pages} {1934--1936},\ \bibinfo {note}
  {https://doi.org/10.18429/JACoW-IPAC2018-WEPAF051}\BibitemShut {NoStop}%
\bibitem [{\citenamefont {{ALEGRO Collaboration}}(2019)}]{alegro-2019-a}%
  \BibitemOpen
  \bibfield  {author} {\bibinfo {author} {\bibnamefont {{ALEGRO
  Collaboration}}},\ }\href {https://doi.org/10.48550/ARXIV.1901.10370}
  {\bibinfo {title} {Towards an advanced linear international collider}}
  (\bibinfo {year} {2019})\BibitemShut {NoStop}%
\bibitem [{\citenamefont {Sinclair}\ \emph {et~al.}(2007)\citenamefont
  {Sinclair}, \citenamefont {Adderley}, \citenamefont {Dunham}, \citenamefont
  {Hansknecht}, \citenamefont {Hartmann}, \citenamefont {Poelker},
  \citenamefont {Price}, \citenamefont {Rutt}, \citenamefont {Schneider},\ and\
  \citenamefont {Steigerwald}}]{sinclair-2007-a}%
  \BibitemOpen
  \bibfield  {author} {\bibinfo {author} {\bibfnamefont {C.~K.}\ \bibnamefont
  {Sinclair}}, \bibinfo {author} {\bibfnamefont {P.~A.}\ \bibnamefont
  {Adderley}}, \bibinfo {author} {\bibfnamefont {B.~M.}\ \bibnamefont
  {Dunham}}, \bibinfo {author} {\bibfnamefont {J.~C.}\ \bibnamefont
  {Hansknecht}}, \bibinfo {author} {\bibfnamefont {P.}~\bibnamefont
  {Hartmann}}, \bibinfo {author} {\bibfnamefont {M.}~\bibnamefont {Poelker}},
  \bibinfo {author} {\bibfnamefont {J.~S.}\ \bibnamefont {Price}}, \bibinfo
  {author} {\bibfnamefont {P.~M.}\ \bibnamefont {Rutt}}, \bibinfo {author}
  {\bibfnamefont {W.~J.}\ \bibnamefont {Schneider}},\ and\ \bibinfo {author}
  {\bibfnamefont {M.}~\bibnamefont {Steigerwald}},\ }\bibfield  {title}
  {\bibinfo {title} {Development of a high average current polarized electron
  source with long cathode operational lifetime},\ }\href
  {https://doi.org/10.1103/PhysRevSTAB.10.023501} {\bibfield  {journal}
  {\bibinfo  {journal} {Phys. Rev. ST Accel. Beams}\ }\textbf {\bibinfo
  {volume} {10}},\ \bibinfo {pages} {023501} (\bibinfo {year}
  {2007})}\BibitemShut {NoStop}%
\bibitem [{\citenamefont {Aulenbacher}\ \emph {et~al.}(1996)\citenamefont
  {Aulenbacher}, \citenamefont {Suberlucq}, \citenamefont {Tang}, \citenamefont
  {Sheppard}, \citenamefont {Mulhollan}, \citenamefont {Delahaye},
  \citenamefont {Madsen}, \citenamefont {Clendenin}, \citenamefont {Bossart},\
  and\ \citenamefont {Braun}}]{aulenbacher-1996-a}%
  \BibitemOpen
  \bibfield  {author} {\bibinfo {author} {\bibfnamefont {K.}~\bibnamefont
  {Aulenbacher}}, \bibinfo {author} {\bibfnamefont {G.}~\bibnamefont
  {Suberlucq}}, \bibinfo {author} {\bibfnamefont {H.}~\bibnamefont {Tang}},
  \bibinfo {author} {\bibfnamefont {J.}~\bibnamefont {Sheppard}}, \bibinfo
  {author} {\bibfnamefont {G.~A.}\ \bibnamefont {Mulhollan}}, \bibinfo {author}
  {\bibfnamefont {J.~P.}\ \bibnamefont {Delahaye}}, \bibinfo {author}
  {\bibfnamefont {J.}~\bibnamefont {Madsen}}, \bibinfo {author} {\bibfnamefont
  {J.~E.}\ \bibnamefont {Clendenin}}, \bibinfo {author} {\bibfnamefont
  {R.}~\bibnamefont {Bossart}},\ and\ \bibinfo {author} {\bibfnamefont
  {H.}~\bibnamefont {Braun}},\ }\href {https://cds.cern.ch/record/307235}
  {\emph {\bibinfo {title} {{RF guns and the production of polarized
  electrons}}}},\ \bibinfo {type} {Tech. Rep.}\ (\bibinfo  {institution}
  {CERN},\ \bibinfo {address} {Geneva},\ \bibinfo {year} {1996})\BibitemShut
  {NoStop}%
\bibitem [{\citenamefont {Aleksandrov}\ \emph {et~al.}(1998)\citenamefont
  {Aleksandrov}, \citenamefont {Konstntinov}, \citenamefont {Logatchev},
  \citenamefont {Starostenko},\ and\ \citenamefont
  {Novokhatskii}}]{aleksandrov-1998-a}%
  \BibitemOpen
  \bibfield  {author} {\bibinfo {author} {\bibfnamefont {A.~V.}\ \bibnamefont
  {Aleksandrov}}, \bibinfo {author} {\bibfnamefont {E.~S.}\ \bibnamefont
  {Konstntinov}}, \bibinfo {author} {\bibfnamefont {P.~V.}\ \bibnamefont
  {Logatchev}}, \bibinfo {author} {\bibfnamefont {A.~A.}\ \bibnamefont
  {Starostenko}},\ and\ \bibinfo {author} {\bibfnamefont {A.~V.}\ \bibnamefont
  {Novokhatskii}},\ }\bibfield  {title} {\bibinfo {title} {{High Power Test of
  GaAs Photocathode in RF Gun}},\ }in\ \href
  {https://cds.cern.ch/record/859144} {\emph {\bibinfo {booktitle} {{ 6th
  European Particle Accelerator Conference, Stockholm, Sweden (EPAC98)}}}}\
  (\bibinfo {year} {1998})\ pp.\ \bibinfo {pages} {1450--1452}\BibitemShut
  {NoStop}%
\bibitem [{\citenamefont {Fliller}\ \emph {et~al.}(2005)\citenamefont
  {Fliller}, \citenamefont {Anderson}, \citenamefont {Edwards}, \citenamefont
  {Bluem}, \citenamefont {Schultheiss}, \citenamefont {Sinclair},\ and\
  \citenamefont {Huening}}]{fliller-2005-a}%
  \BibitemOpen
  \bibfield  {author} {\bibinfo {author} {\bibfnamefont {R.}~\bibnamefont
  {Fliller}}, \bibinfo {author} {\bibfnamefont {T.}~\bibnamefont {Anderson}},
  \bibinfo {author} {\bibfnamefont {H.}~\bibnamefont {Edwards}}, \bibinfo
  {author} {\bibfnamefont {H.}~\bibnamefont {Bluem}}, \bibinfo {author}
  {\bibfnamefont {T.}~\bibnamefont {Schultheiss}}, \bibinfo {author}
  {\bibfnamefont {C.}~\bibnamefont {Sinclair}},\ and\ \bibinfo {author}
  {\bibfnamefont {M.}~\bibnamefont {Huening}},\ }\bibfield  {title} {\bibinfo
  {title} {Progress on using {NEA} cathodes in an rf gun},\ }in\ \href
  {https://doi.org/10.1109/PAC.2005.1591236} {\emph {\bibinfo {booktitle}
  {Proceedings of the 2005 Particle Accelerator Conference}}}\ (\bibinfo {year}
  {2005})\ pp.\ \bibinfo {pages} {2708--2710}\BibitemShut {NoStop}%
\bibitem [{\citenamefont {Bargmann}\ \emph {et~al.}(1959)\citenamefont
  {Bargmann}, \citenamefont {Michel},\ and\ \citenamefont
  {Telegdi}}]{bargmann-1959-a}%
  \BibitemOpen
  \bibfield  {author} {\bibinfo {author} {\bibfnamefont {V.}~\bibnamefont
  {Bargmann}}, \bibinfo {author} {\bibfnamefont {L.}~\bibnamefont {Michel}},\
  and\ \bibinfo {author} {\bibfnamefont {V.~L.}\ \bibnamefont {Telegdi}},\
  }\bibfield  {title} {\bibinfo {title} {Precession of the polarization of
  particles moving in a homogeneous electromagnetic field},\ }\href
  {https://doi.org/10.1103/PhysRevLett.2.435} {\bibfield  {journal} {\bibinfo
  {journal} {Phys. Rev. Lett.}\ }\textbf {\bibinfo {volume} {2}},\ \bibinfo
  {pages} {435} (\bibinfo {year} {1959})}\BibitemShut {NoStop}%
\bibitem [{\citenamefont {Sagan}(2006)}]{bmad2006}%
  \BibitemOpen
  \bibfield  {author} {\bibinfo {author} {\bibfnamefont {D.}~\bibnamefont
  {Sagan}},\ }\bibfield  {title} {\bibinfo {title} {{{BMAD}: A relativistic
  charged particle simulation library}},\ }\bibfield  {booktitle} {\emph
  {\bibinfo {booktitle} {{Computational accelerator physics. Proceedings, 8th
  International Conference, ICAP 2004, St. Petersburg, Russia, June 29-July 2,
  2004}}},\ }\href {https://doi.org/https://doi.org/10.1016/j.nima.2005.11.001}
  {\bibfield  {journal} {\bibinfo  {journal} {Nucl. Instrum. Meth.}\ }\textbf
  {\bibinfo {volume} {A558}},\ \bibinfo {pages} {356} (\bibinfo {year}
  {2006})},\ \bibinfo {note} {proceedings of the 8th International
  Computational Accelerator Physics Conference}\BibitemShut {NoStop}%
\bibitem [{\citenamefont {Yakimenko}\ \emph {et~al.}(2019)\citenamefont
  {Yakimenko}, \citenamefont {Meuren}, \citenamefont {Del~Gaudio},
  \citenamefont {Baumann}, \citenamefont {Fedotov}, \citenamefont {Fiuza},
  \citenamefont {Grismayer}, \citenamefont {Hogan}, \citenamefont {Pukhov},
  \citenamefont {Silva},\ and\ \citenamefont {White}}]{yakimenko-2019-a}%
  \BibitemOpen
  \bibfield  {author} {\bibinfo {author} {\bibfnamefont {V.}~\bibnamefont
  {Yakimenko}}, \bibinfo {author} {\bibfnamefont {S.}~\bibnamefont {Meuren}},
  \bibinfo {author} {\bibfnamefont {F.}~\bibnamefont {Del~Gaudio}}, \bibinfo
  {author} {\bibfnamefont {C.}~\bibnamefont {Baumann}}, \bibinfo {author}
  {\bibfnamefont {A.}~\bibnamefont {Fedotov}}, \bibinfo {author} {\bibfnamefont
  {F.}~\bibnamefont {Fiuza}}, \bibinfo {author} {\bibfnamefont
  {T.}~\bibnamefont {Grismayer}}, \bibinfo {author} {\bibfnamefont {M.~J.}\
  \bibnamefont {Hogan}}, \bibinfo {author} {\bibfnamefont {A.}~\bibnamefont
  {Pukhov}}, \bibinfo {author} {\bibfnamefont {L.~O.}\ \bibnamefont {Silva}},\
  and\ \bibinfo {author} {\bibfnamefont {G.}~\bibnamefont {White}},\ }\bibfield
   {title} {\bibinfo {title} {Prospect of studying nonperturbative {QED} with
  beam-beam collisions},\ }\href
  {https://doi.org/10.1103/PhysRevLett.122.190404} {\bibfield  {journal}
  {\bibinfo  {journal} {Phys. Rev. Lett.}\ }\textbf {\bibinfo {volume} {122}},\
  \bibinfo {pages} {190404} (\bibinfo {year} {2019})}\BibitemShut {NoStop}%
\bibitem [{\citenamefont {Latina}\ and\ \citenamefont
  {Solyak}(2010)}]{latina-2010-a}%
  \BibitemOpen
  \bibfield  {author} {\bibinfo {author} {\bibfnamefont {A.}~\bibnamefont
  {Latina}}\ and\ \bibinfo {author} {\bibfnamefont {N.}~\bibnamefont
  {Solyak}},\ }\bibfield  {title} {\bibinfo {title} {{Single-Stage Bunch
  Compressor for ILC-SB2009}},\ }\href
  {https://accelconf.web.cern.ch/IPAC10/papers/thpe042.pdf} {\bibfield
  {journal} {\bibinfo  {journal} {Conf. Proc. C}\ }\textbf {\bibinfo {volume}
  {100523}},\ \bibinfo {pages} {THPE042} (\bibinfo {year} {2010})}\BibitemShut
  {NoStop}%
\bibitem [{\citenamefont {Hessami}\ and\ \citenamefont
  {Gessner}(2022)}]{hessami-2022-a}%
  \BibitemOpen
  \bibfield  {author} {\bibinfo {author} {\bibfnamefont {R.}~\bibnamefont
  {Hessami}}\ and\ \bibinfo {author} {\bibfnamefont {S.}~\bibnamefont
  {Gessner}},\ }\href@noop {} {\emph {\bibinfo {title} {{A Compact Source of
  Positron Beams with Small Thermal Emittance}}}},\ \bibinfo {type} {Tech.
  Rep.}\ (\bibinfo  {institution} {SLAC},\ \bibinfo {address} {Menlo Park,
  CA},\ \bibinfo {year} {2022})\ \bibinfo {note} {to be published (private
  communication with S. Gessner)}\BibitemShut {NoStop}%
\bibitem [{\citenamefont {Abbott}\ \emph {et~al.}(2016)\citenamefont {Abbott},
  \citenamefont {Adderley}, \citenamefont {Adeyemi}, \citenamefont {Aguilera},
  \citenamefont {Ali}, \citenamefont {Areti}, \citenamefont {Baylac},
  \citenamefont {Benesch}, \citenamefont {Bosson}, \citenamefont {Cade},
  \citenamefont {Camsonne}, \citenamefont {Cardman}, \citenamefont {Clark},
  \citenamefont {Cole}, \citenamefont {Covert}, \citenamefont {Cuevas},
  \citenamefont {Dadoun}, \citenamefont {Dale}, \citenamefont {Dong},
  \citenamefont {Dumas}, \citenamefont {Fanchini}, \citenamefont {Forest},
  \citenamefont {Forman}, \citenamefont {Freyberger}, \citenamefont
  {Froidefond}, \citenamefont {Golge}, \citenamefont {Grames}, \citenamefont
  {Gu\`eye}, \citenamefont {Hansknecht}, \citenamefont {Harrell}, \citenamefont
  {Hoskins}, \citenamefont {Hyde}, \citenamefont {Josey}, \citenamefont
  {Kazimi}, \citenamefont {Kim}, \citenamefont {Machie}, \citenamefont
  {Mahoney}, \citenamefont {Mammei}, \citenamefont {Marton}, \citenamefont
  {McCarter}, \citenamefont {McCaughan}, \citenamefont {McHugh}, \citenamefont
  {McNulty}, \citenamefont {Mesick}, \citenamefont {Michaelides}, \citenamefont
  {Michaels}, \citenamefont {Moffit}, \citenamefont {Moser}, \citenamefont
  {Mu\~noz Camacho}, \citenamefont {Muraz}, \citenamefont {Opper},
  \citenamefont {Poelker}, \citenamefont {R\'eal}, \citenamefont {Richardson},
  \citenamefont {Setiniyaz}, \citenamefont {Stutzman}, \citenamefont
  {Suleiman}, \citenamefont {Tennant}, \citenamefont {Tsai}, \citenamefont
  {Turner}, \citenamefont {Ungaro}, \citenamefont {Variola}, \citenamefont
  {Voutier}, \citenamefont {Wang},\ and\ \citenamefont
  {Zhang}}]{abbott-2016-a}%
  \BibitemOpen
  \bibfield  {author} {\bibinfo {author} {\bibfnamefont {D.}~\bibnamefont
  {Abbott}}, \bibinfo {author} {\bibfnamefont {P.}~\bibnamefont {Adderley}},
  \bibinfo {author} {\bibfnamefont {A.}~\bibnamefont {Adeyemi}}, \bibinfo
  {author} {\bibfnamefont {P.}~\bibnamefont {Aguilera}}, \bibinfo {author}
  {\bibfnamefont {M.}~\bibnamefont {Ali}}, \bibinfo {author} {\bibfnamefont
  {H.}~\bibnamefont {Areti}}, \bibinfo {author} {\bibfnamefont
  {M.}~\bibnamefont {Baylac}}, \bibinfo {author} {\bibfnamefont
  {J.}~\bibnamefont {Benesch}}, \bibinfo {author} {\bibfnamefont
  {G.}~\bibnamefont {Bosson}}, \bibinfo {author} {\bibfnamefont
  {B.}~\bibnamefont {Cade}}, \bibinfo {author} {\bibfnamefont {A.}~\bibnamefont
  {Camsonne}}, \bibinfo {author} {\bibfnamefont {L.~S.}\ \bibnamefont
  {Cardman}}, \bibinfo {author} {\bibfnamefont {J.}~\bibnamefont {Clark}},
  \bibinfo {author} {\bibfnamefont {P.}~\bibnamefont {Cole}}, \bibinfo {author}
  {\bibfnamefont {S.}~\bibnamefont {Covert}}, \bibinfo {author} {\bibfnamefont
  {C.}~\bibnamefont {Cuevas}}, \bibinfo {author} {\bibfnamefont
  {O.}~\bibnamefont {Dadoun}}, \bibinfo {author} {\bibfnamefont
  {D.}~\bibnamefont {Dale}}, \bibinfo {author} {\bibfnamefont {H.}~\bibnamefont
  {Dong}}, \bibinfo {author} {\bibfnamefont {J.}~\bibnamefont {Dumas}},
  \bibinfo {author} {\bibfnamefont {E.}~\bibnamefont {Fanchini}}, \bibinfo
  {author} {\bibfnamefont {T.}~\bibnamefont {Forest}}, \bibinfo {author}
  {\bibfnamefont {E.}~\bibnamefont {Forman}}, \bibinfo {author} {\bibfnamefont
  {A.}~\bibnamefont {Freyberger}}, \bibinfo {author} {\bibfnamefont
  {E.}~\bibnamefont {Froidefond}}, \bibinfo {author} {\bibfnamefont
  {S.}~\bibnamefont {Golge}}, \bibinfo {author} {\bibfnamefont
  {J.}~\bibnamefont {Grames}}, \bibinfo {author} {\bibfnamefont
  {P.}~\bibnamefont {Gu\`eye}}, \bibinfo {author} {\bibfnamefont
  {J.}~\bibnamefont {Hansknecht}}, \bibinfo {author} {\bibfnamefont
  {P.}~\bibnamefont {Harrell}}, \bibinfo {author} {\bibfnamefont
  {J.}~\bibnamefont {Hoskins}}, \bibinfo {author} {\bibfnamefont
  {C.}~\bibnamefont {Hyde}}, \bibinfo {author} {\bibfnamefont {B.}~\bibnamefont
  {Josey}}, \bibinfo {author} {\bibfnamefont {R.}~\bibnamefont {Kazimi}},
  \bibinfo {author} {\bibfnamefont {Y.}~\bibnamefont {Kim}}, \bibinfo {author}
  {\bibfnamefont {D.}~\bibnamefont {Machie}}, \bibinfo {author} {\bibfnamefont
  {K.}~\bibnamefont {Mahoney}}, \bibinfo {author} {\bibfnamefont
  {R.}~\bibnamefont {Mammei}}, \bibinfo {author} {\bibfnamefont
  {M.}~\bibnamefont {Marton}}, \bibinfo {author} {\bibfnamefont
  {J.}~\bibnamefont {McCarter}}, \bibinfo {author} {\bibfnamefont
  {M.}~\bibnamefont {McCaughan}}, \bibinfo {author} {\bibfnamefont
  {M.}~\bibnamefont {McHugh}}, \bibinfo {author} {\bibfnamefont
  {D.}~\bibnamefont {McNulty}}, \bibinfo {author} {\bibfnamefont {K.~E.}\
  \bibnamefont {Mesick}}, \bibinfo {author} {\bibfnamefont {T.}~\bibnamefont
  {Michaelides}}, \bibinfo {author} {\bibfnamefont {R.}~\bibnamefont
  {Michaels}}, \bibinfo {author} {\bibfnamefont {B.}~\bibnamefont {Moffit}},
  \bibinfo {author} {\bibfnamefont {D.}~\bibnamefont {Moser}}, \bibinfo
  {author} {\bibfnamefont {C.}~\bibnamefont {Mu\~noz Camacho}}, \bibinfo
  {author} {\bibfnamefont {J.-F.}\ \bibnamefont {Muraz}}, \bibinfo {author}
  {\bibfnamefont {A.}~\bibnamefont {Opper}}, \bibinfo {author} {\bibfnamefont
  {M.}~\bibnamefont {Poelker}}, \bibinfo {author} {\bibfnamefont {J.-S.}\
  \bibnamefont {R\'eal}}, \bibinfo {author} {\bibfnamefont {L.}~\bibnamefont
  {Richardson}}, \bibinfo {author} {\bibfnamefont {S.}~\bibnamefont
  {Setiniyaz}}, \bibinfo {author} {\bibfnamefont {M.}~\bibnamefont {Stutzman}},
  \bibinfo {author} {\bibfnamefont {R.}~\bibnamefont {Suleiman}}, \bibinfo
  {author} {\bibfnamefont {C.}~\bibnamefont {Tennant}}, \bibinfo {author}
  {\bibfnamefont {C.}~\bibnamefont {Tsai}}, \bibinfo {author} {\bibfnamefont
  {D.}~\bibnamefont {Turner}}, \bibinfo {author} {\bibfnamefont
  {M.}~\bibnamefont {Ungaro}}, \bibinfo {author} {\bibfnamefont
  {A.}~\bibnamefont {Variola}}, \bibinfo {author} {\bibfnamefont
  {E.}~\bibnamefont {Voutier}}, \bibinfo {author} {\bibfnamefont
  {Y.}~\bibnamefont {Wang}},\ and\ \bibinfo {author} {\bibfnamefont
  {Y.}~\bibnamefont {Zhang}} (\bibinfo {collaboration} {PEPPo Collaboration}),\
  }\bibfield  {title} {\bibinfo {title} {Production of highly polarized
  positrons using polarized electrons at mev energies},\ }\href
  {https://doi.org/10.1103/PhysRevLett.116.214801} {\bibfield  {journal}
  {\bibinfo  {journal} {Phys. Rev. Lett.}\ }\textbf {\bibinfo {volume} {116}},\
  \bibinfo {pages} {214801} (\bibinfo {year} {2016})}\BibitemShut {NoStop}%
\bibitem [{\citenamefont {Kuriki}\ \emph {et~al.}(2018)\citenamefont {Kuriki}
  \emph {et~al.}}]{kuriki-2018-a}%
  \BibitemOpen
  \bibfield  {author} {\bibinfo {author} {\bibfnamefont {M.}~\bibnamefont
  {Kuriki}} \emph {et~al.},\ }\bibfield  {title} {\bibinfo {title} {{High
  Aspect Ratio Beam Generation with the Phase-space Rotation Technique for
  Linear Colliders}},\ }in\ \href
  {https://doi.org/10.18429/JACoW-LINAC2018-THPO005} {\emph {\bibinfo
  {booktitle} {{29th International Linear Accelerator Conference}}}}\ (\bibinfo
  {year} {2018})\ p.\ \bibinfo {pages} {THPO005}\BibitemShut {NoStop}%
\end{thebibliography}%

\end{document}